\documentclass[longauth]{aa}

\usepackage{graphicx}
\usepackage{txfonts}
\usepackage{url}
\usepackage{hyperref}
\usepackage[separate-uncertainty=true,per-mode=symbol-or-fraction]{siunitx}
\DeclareSIUnit{\AU}{AU}
\DeclareSIUnit{\solarradius}{R_\sun}

\newcommand{\maxt}{{\text{max}}}

\usepackage[dvipsnames]{xcolor}
\usepackage{tikz}
\definecolor{red}{HTML}{d62728}
\definecolor{orange}{HTML}{ff7f0e}
\definecolor{green}{HTML}{2ca02c}
\definecolor{blue}{HTML}{1f77b4}
\definecolor{purple}{HTML}{9467bd}
\definecolor{light-gray}{gray}{0.9}
\usetikzlibrary{calc}

\begin{document}

\title{Radial Evolution of the April 2020 Stealth\\ Coronal Mass Ejection between 0.8 and 1 AU}
\subtitle{A Comparison of Forbush Decreases at Solar Orbiter and Earth}

\author{
    Johan L. Freiherr von Forstner\inst{\ref{inst:cau}} \and
    Mateja Dumbovi\'{c}\inst{\ref{inst:hvar_obs}} \and
    Christian Möstl\inst{\ref{inst:iwf_graz}} \and
    Jingnan Guo\inst{\ref{inst:hefei},\ref{inst:hefei1},\ref{inst:cau}} \and
    Athanasios Papaioannou\inst{\ref{inst:athens}}
    Robert Elftmann\inst{\ref{inst:cau}} \and
    Zigong Xu\inst{\ref{inst:cau}} \and
    Jan Christoph Terasa\inst{\ref{inst:cau}} \and
    Alexander Kollhoff\inst{\ref{inst:cau}} \and
    Robert F. Wimmer-Schweingruber\inst{\ref{inst:cau}} \and
    Javier Rodr\'{i}guez-Pacheco\inst{\ref{inst:alcala}} \and
    Andreas J. Weiss\inst{\ref{inst:iwf_graz}} \and
    Jürgen Hinterreiter\inst{\ref{inst:iwf_graz}} \and
    Tanja Amerstorfer\inst{\ref{inst:iwf_graz}} \and
    Maike Bauer\inst{\ref{inst:iwf_graz}} \and
    Anatoly V. Belov\inst{\ref{inst:izmiran}} \and
    Maria A. Abunina\inst{\ref{inst:izmiran}} \and
    Timothy Horbury\inst{\ref{inst:imperial}} \and
    Emma E. Davies\inst{\ref{inst:imperial}} \and
    Helen O'Brien\inst{\ref{inst:imperial}} \and
    Robert C. Allen \inst{\ref{inst:apl}} \and
    G. Bruce Andrews \inst{\ref{inst:apl}} \and
    Lars Berger\inst{\ref{inst:cau}} \and
    Sebastian Boden\inst{\ref{inst:cau},\ref{inst:dsi_boden}} \and
    Ignacio Cernuda Cangas\inst{\ref{inst:alcala}} \and
    Sandra Eldrum\inst{\ref{inst:cau}} \and
    Francisco Espinosa Lara\inst{\ref{inst:alcala}} \and
    Raúl Gómez Herrero\inst{\ref{inst:alcala}} \and
    John R. Hayes \inst{\ref{inst:apl}} \and
    George C. Ho\inst{\ref{inst:apl}} \and
    Shrinivasrao R. Kulkarni\inst{\ref{inst:cau},\ref{inst:desy_shri}} \and
    W. Jeffrey Lees \inst{\ref{inst:apl}} \and
    César Martín\inst{\ref{inst:cau},\ref{inst:dlr}} \and
    Glenn M. Mason\inst{\ref{inst:apl}} \and
    Daniel Pacheco\inst{\ref{inst:cau}} \and
    Manuel Prieto Mateo\inst{\ref{inst:alcala}} \and
    Ali Ravanbakhsh\inst{\ref{inst:cau},\ref{inst:mpi}} \and
    Oscar Rodríguez Polo\inst{\ref{inst:alcala}} \and
    Sebastián Sánchez Prieto \inst{\ref{inst:alcala}} \and
    Charles E. Schlemm \inst{\ref{inst:apl}} \and
    Helmut Seifert \inst{\ref{inst:apl}} \and
    Kush Tyagi \inst{\ref{inst:lasp}} \and
    Mahesh Yedla\inst{\ref{inst:cau},\ref{inst:mpi}}
}

\institute{Institut für Experimentelle und Angewandte Physik, Christian-Albrechts-Universität zu Kiel, 24098 Kiel, Germany\\
    \email{forstner@physik.uni-kiel.de}\label{inst:cau} \and
    Hvar Observatory, Faculty of Geodesy, University of Zagreb, Zagreb, Croatia\label{inst:hvar_obs} \and
    Space Research Institute, Austrian Academy of Sciences, Graz, Austria\label{inst:iwf_graz} \and
    School of Earth and Space Sciences, University of Science and Technology of China, Hefei, China\label{inst:hefei} \and
    CAS Center for Excellence in Comparative Planetology, Hefei, PR China\label{inst:hefei1}
    \and
    Institute for Astronomy, Astrophysics, Space Applications and Remote Sensing (IAASARS), National Observatory of Athens, Athens, Greece\label{inst:athens} \and
    Space Research Group, Universidad de Alcalá, 28805 Alcalá de Henares, Spain\label{inst:alcala} \and
    Pushkov Institute of Terrestrial Magnetism, Ionosphere, and Radio Wave Propagation, Russian Academy of Sciences (IZMIRAN), Troitsk, Moscow, Russia\label{inst:izmiran} \and
    Imperial College London, London SW7 2AZ, United Kingdom\label{inst:imperial} \and
    Johns Hopkins University Applied Physics Laboratory, Laurel, MD, United States \label{inst:apl} \and
    Now at: DSI Datensicherheit GmbH, Rodendamm 34, 28816 Stuhr\label{inst:dsi_boden} \and
    Now at: Deutsches Elektronen-Synchrotron (DESY), Platanenallee 6, 15738 Zeuthen, Germany\label{inst:desy_shri} \and
    Now at: Department of Extrasolar Planets and Atmospheres, German Aerospace Center (DLR), Berlin, Germany\label{inst:dlr} \and
    Now at: Max-Planck-Institute for Solar System Research, Justus-von-Liebig-Weg 3, 37077 Göttingen, Germany\label{inst:mpi} \and
    Laboratory for Atmospheric and Space Physics, University of Colorado Boulder, Boulder, CO, USA\label{inst:lasp}
}

\abstract
{}
{We present observations of the first coronal mass ejection (CME) observed at the Solar Orbiter spacecraft on April 19, 2020, and the associated Forbush decrease (FD) measured by its High Energy Telescope (HET). This CME is a multispacecraft event also seen near Earth the next day.}
{We highlight the capabilities of HET for observing small short-term variations of the galactic cosmic ray count rate using its single detector counters. The analytical ForbMod model is applied to the FD measurements to reproduce the Forbush decrease at both locations. Input parameters for the model are derived from both in situ and remote-sensing observations of the CME.}
{The very slow ($\sim\SI{350}{\kilo\meter\per\second}$) stealth CME caused a FD with an amplitude of \SI{3}{\percent}  in the low-energy cosmic ray measurements at HET and \SI{2}{\percent} in a comparable channel of the Cosmic Ray Telescope for the Effects of Radiation (CRaTER) on the Lunar Reconnaissance Orbiter, as well as a \SI{1}{\percent} decrease in neutron monitor measurements. Significant differences are observed in the expansion behavior of the CME at different locations, which may be related to influence of the following high speed solar wind stream. Under certain assumptions, ForbMod is able to reproduce the observed FDs in low-energy cosmic ray measurements from HET as well as CRaTER, but with the same input parameters, the results do not agree with the FD amplitudes at higher energies measured by neutron monitors on Earth. We study these discrepancies and provide possible explanations.}
{This study highlights that the novel measurements of the Solar Orbiter can be coordinated with other spacecraft to improve our understanding of space weather in the inner heliosphere. Multi-spacecraft observations combined with data-based modeling are also essential to understand the propagation and evolution of CMEs as well as their space weather impacts.}


\maketitle

\section{Introduction}

\begin{figure}
	\centering
	\includegraphics[width=\hsize]{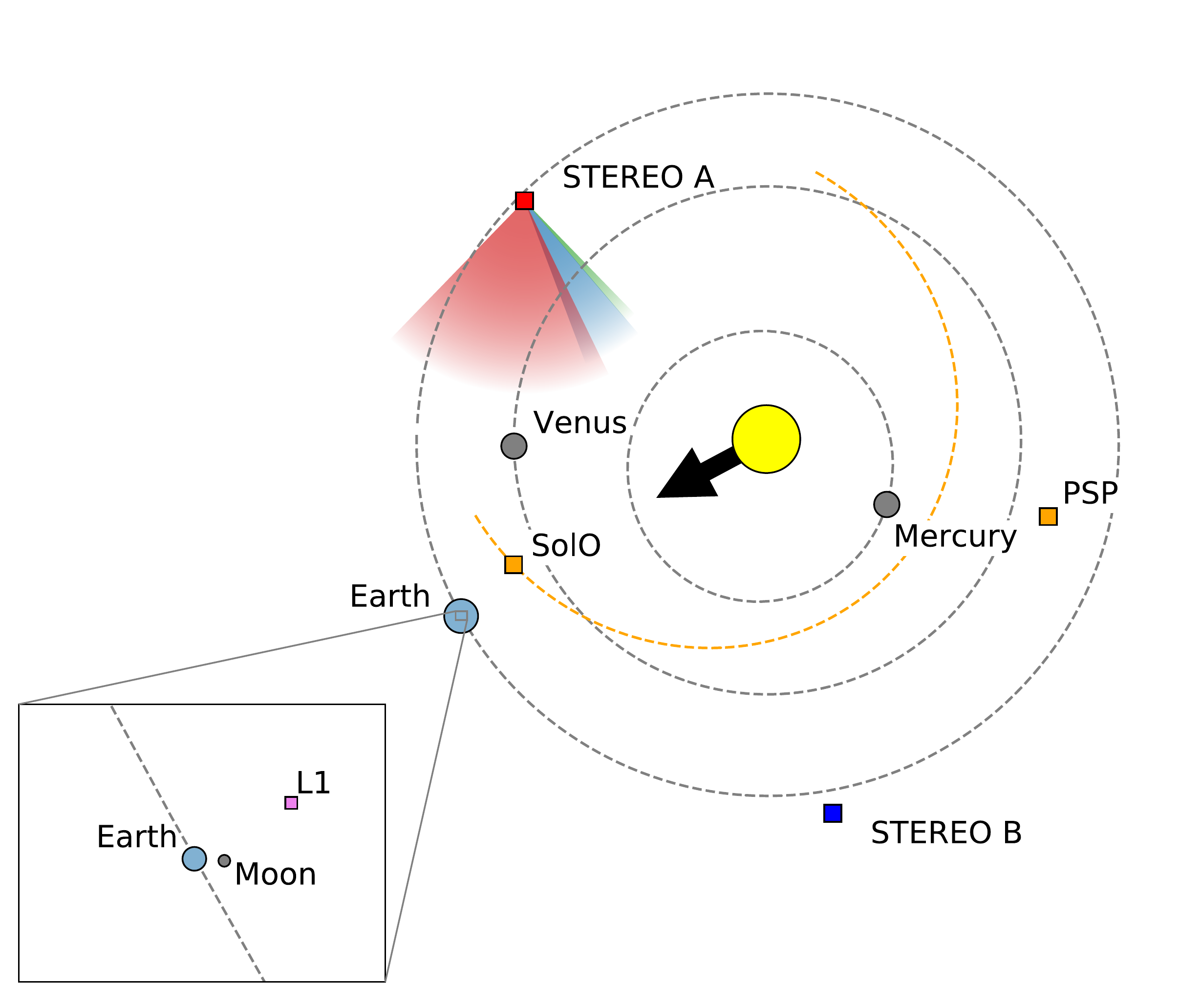}
	\caption{Locations of planets and spacecraft in the inner solar system on April 20, 2020, the day the CME 
		arrived at Earth. The trajectory of Solar Orbiter (SolO) is shown as an orange dashed line, and PSP denotes the 
		location of Parker Solar Probe. The large black arrow describes the approximate propagation direction of the CME, and the colored segments next to STEREO-A show the fields of view of the remote sensing instruments COR1/COR2 (green), HI1 (blue) and HI2 (red). The inset shows a zoomed-in view of the relative positions of Earth, the Moon and the Lagrange point L1, where the Wind spacecraft is located.}
	\label{fig:solar_system}
\end{figure}

On April 19, 2020, a coronal mass ejection (CME) passed the Solar Orbiter \citep[SolO,][]{Mueller-2020-SolO} spacecraft -- the first large-scale 
flux rope CME seen in situ at SolO. At this time, the spacecraft was closely aligned in heliospheric longitude with Earth (less than \SI{4}{\degree} separation), and it was located at a radial distance of 0.8 AU from the Sun, as shown in Fig. \ref{fig:solar_system}. Consequently, the same 
slow CME ($v <$ 400 km/s) was also observed near Earth the next day, causing the first geomagnetic storm of the year with a Dst index of \SI{-59}{\nano\tesla} and Kp index of 5. During the event, SolO was still in its Near Earth 
Commissioning Phase, which ended on June 15, 2020, but nevertheless, some of the in situ instruments, including the Energetic Particle Detector suite \citep[EPD,][]{RodriguezPacheco-2019-EPD} and the magnetometer \citep[MAG,][]{Horbury-2020-MAG} were already taking continuous measurements and were able to observe signatures of the CME.
In addition, the STEREO-A spacecraft had a sufficient longitudinal separation of $\sim\SI{75}{\degree}$ from SolO and the Earth and thus was able to provide excellent remote sensing observations of the CME propagation from a side view.
This event observed at both SolO and Earth provides an excellent example for the coordinated science that is possible with SolO and other heliophysics missions in the solar system.

CMEs, clouds of magnetized plasma ejected from the Sun, are one of the key phenomena in space weather research, as they can cause severe geomagnetic storms \citep{Kilpua2017} disrupting terrestrial infrastructure. The shocks driven by CMEs are also partly responsible for energetic particles in the heliosphere \citep{Reames-2013}, which may pose radiation danger to astronauts and spacecraft. Consequently, two of the four main scientific questions of the Solar Orbiter mission \citep{Mueller-2013-SolO} are also linked to the better understanding of CMEs.

Forbush decreases (FDs), first observed by and later named after Scott E. \citet{Forbush-1937}, are short-term decreases of the galactic cosmic ray (GCR) flux, caused by the passage of magnetic field structures in the solar wind, such as CMEs or stream interaction regions (SIRs). Such magnetic structures can act as a barrier for the propagation of GCRs, e.g. because the GCRs need to diffuse across a strong field, so that the observed flux is temporarily decreased at the locations these structures pass.  The decrease phase usually takes less than 1 day, followed by an often slower recovery to the previous level (on the order of 1 week). In the case of CMEs, FDs are driven by both the turbulent shock/sheath region (if present) as well as the following magnetic ejecta, two effects which can sometimes be clearly separated when a two-step decrease is observed \citep[e.g.][]{Cane-2000}. The amplitude of a FD depends not only on the properties of the heliospheric structure, but also on the energy of the observed GCR particles: lower energy particles are modulated more easily and thus typically show larger FDs \citep[e.g.][]{Lockwood1971,Lockwood1991,Cane-2000,Guo-2020-new}.
In the past, the study of FDs was mainly based on data from neutron monitors on the surface of the Earth, but nowadays, GCR measurements suitable for FD studies are also available from many spacecraft in the near-Earth space as well as on other solar system bodies, and these have been routinely used for multi-spacecraft studies \citep[e.g.][]{Cane1994,Lockwood1991,Forstner-2018,Forstner-2019,Forstner-2020,Witasse-2017,Winslow-2018}. In all cases, it is important to take into account the energy dependence of the FD amplitude, as such instruments may be sensitive to different GCR energies.

In this work, we present the EPD observations of the FD associated with the April 19 CME at SolO, as well as the corresponding observations at Earth. We describe which EPD data products are best suited to make measurements of FDs, and we analyze these data to see how the CME affected the GCR flux at different heliospheric locations and at different particle energies.
We also employ the ForbMod model to reproduce the observed FD and gain insight into the how the large-scale evolution of the CME structure affected the properties of the FD.
A study by \citet{Davies-2021} complements this work by investigating the magnetic field measurements at both Solar Orbiter and Earth in more detail.
In Sect. \ref{sec:data}, we will introduce the different instruments used as data sources in this study, followed by an overview of our modeling methods in Sect. \ref{sec:models}. The measurement and modeling results will then be presented in Sect. \ref{sec:results}, and discussed in more detail in Sect. \ref{sec:discussion_conclusions}.

\section{Data sources}
\label{sec:data}

\subsection{HET on Solar Orbiter}
\label{subsec:het}

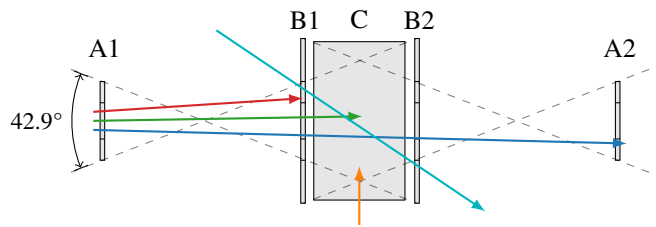
\begin{figure}
    \centering



\begin{tikzpicture}[scale=0.6]
    \tikzset{%
        add/.style args={#1 and #2}{
            to path={%
                ($(\tikztostart)!-#1!(\tikztotarget)$)--($(\tikztotarget)!-#2!(\tikztostart)$)%
                \tikztonodes},add/.default={.2 and .2}}
    }   
    
    \def\Sithickness{0.05}
    
    \draw[fill=light-gray] (-\Sithickness, -0.87) rectangle (\Sithickness, 0.87) node[above,yshift=2mm]{A1};
    \draw (-\Sithickness, -0.4) -- (\Sithickness, -0.4);
    \draw (-\Sithickness, 0.4) -- (\Sithickness, 0.4);
    
    \draw[fill=light-gray] (4.4-\Sithickness, -1.82) rectangle (4.4+\Sithickness, 1.82) node[above]{B1};
    \draw (4.4-\Sithickness, -0.4) -- (4.4+\Sithickness, -0.4);
    \draw (4.4-\Sithickness, 0.4) -- (4.4+\Sithickness, 0.4);
    \draw (4.4-\Sithickness, -0.87) -- (4.4+\Sithickness, -0.87);
    \draw (4.4-\Sithickness, 0.87) -- (4.4+\Sithickness, 0.87);
    
    \draw[fill=light-gray] (4.635, -1.75) rectangle ++(2, 3.5);
    \node[above, yshift=0.5mm] at (5.635, 1.75) {C};
    
    \draw[fill=light-gray] (6.9-\Sithickness, -1.82) rectangle (6.9+\Sithickness, 1.82) node[above]{B2};
    \draw (6.9-\Sithickness, -0.4) -- (6.9+\Sithickness, -0.4);
    \draw (6.9-\Sithickness, 0.4) -- (6.9+\Sithickness, 0.4);
    \draw (6.9-\Sithickness, -0.87) -- (6.9+\Sithickness, -0.87);
    \draw (6.9-\Sithickness, 0.87) -- (6.9+\Sithickness, 0.87);
    
    \draw[fill=light-gray] (11.3-\Sithickness, -0.87) rectangle (11.3+\Sithickness, 0.87) node[above,yshift=2mm]{A2};
    \draw (11.3-\Sithickness, -0.4) -- (11.3+\Sithickness, -0.4);
    \draw (11.3-\Sithickness, 0.4) -- (11.3+\Sithickness, 0.4);
    
    \draw[add=0.1 and 0, dashed, opacity=0.5] (0, -0.87) to (6.9, 1.82);
    \draw[add=0.1 and 0, dashed, opacity=0.5] (0, 0.87) to (6.9, -1.82);
    \draw[add=0.1 and 0, dashed, opacity=0.5] (11.3, -0.87) to (4.4, 1.82);
    \draw[add=0.1 and 0, dashed, opacity=0.5] (11.3, 0.87) to (4.4, -1.82);
    
    \draw[|<->|] (0, 0.87) ++ (180-21.45:0.5) arc (180-21.45:180+21.45:2.4+0.48) node[midway, left] {\small\SI{42.9}{\degree}};

    \draw[thick, red, -latex] (-0.2, 0.2) -- (4.4, 0.5);
    \draw[thick, green, -latex] (-0.2, 0) -- (5.7, 0.1);
    \draw[thick, blue, -latex] (-0.2, -0.2) -- (11.5, -0.5);
    \draw[thick, Aquamarine, -latex] (2.5, 2) -- (8.4, -2);
    \draw[thick, orange, -latex] (5.635, -2.3) -- (5.635, -1);
\end{tikzpicture}
    \caption{Schematic diagram of the HET sensor head. Exemplary particle trajectories ending up in different data products are shown by the arrows: \textcolor{red}{stopping in B}, \textcolor{green}{stopping in C}, \textcolor{blue}{penetrating}, \textcolor{Aquamarine}{GCR channel}, \textcolor{orange}{C single counter}. A 3D graphic of the sensor head is shown in \citet[][Fig. 31]{RodriguezPacheco-2019-EPD}.}
    \label{fig:solohet_sensor_schematic}
\end{figure}

As part of the Energetic Particle Detector suite \citep{RodriguezPacheco-2019-EPD} onboard the Solar Orbiter mission 
\citep{Mueller-2013-SolO,Mueller-2020-SolO}, the High Energy Telescope (HET) is a particle telescope covering the 
high-energy end of the solar energetic particle (SEP) spectrum as well as galactic cosmic rays (GCR). Its two 
double-ended telescopes each consist of four thin \SI{300}{\micro\meter} silicon solid state detectors (named the A1, B1, B2, and A2 detectors) 
and the C detector, a \SI{2}{\centi\meter} thick Bi$_4$Ge$_3$O$_{12}$ (BGO) scintillator, in the center. This detector layout is shown in Fig. \ref{fig:solohet_sensor_schematic}. The C detector is read out using two 
photodiodes placed on either side, named C1 and C2. HET is designed to measure the 
fluxes of electrons above \SI{300}{\kilo\electronvolt}, protons above \SI{7}{\mega\electronvolt} as well as heavier 
ions, with one telescope (HET 1) providing the sunward and antisunward viewing directions (parallel to the mean Parker 
spiral angle), and the other (HET 2) being mounted perpendicular to that to measure particles coming from outside the 
ecliptic plane. The telescopes distinguish between particles stopping in one of the B detectors (B1 or B2, e.g. red arrow in Fig. \ref{fig:solohet_sensor_schematic}), particles stopping in the C (green arrow)
detector, and particles penetrating the whole telescope (blue arrow) to achieve a large energy coverage, and use the 
$\mathrm{d}E/\mathrm{d}x$-$E$-method to separate different particle species. This technique has been in use by many previous space-borne charged particle detectors, including the Interplanetary Monitoring Platform-1 mission in the 1960s \citep{McDonald-1964} as well as more recent instruments such as the Mars Science Laboratory Radiation Assessment Detector \citep{Hassler-2012-MSLRAD} and the Chang'E 4 Lunar Lander Neutrons and Dosimetry experiment \citep{Wimmer-2020-LND}. For more details about the application of the $\mathrm{d}E/\mathrm{d}x$-$E$-method in HET, see \citet[Sect. 7.2.5]{RodriguezPacheco-2019-EPD}.

While the nominal data products of HET are optimized for the study of high intensity SEP events by choosing a rather 
small opening angle to achieve a high energy resolution, these data are not optimal for observing short-term variations 
of the GCR background due to their low level of counting statistics. Alternatively, HET provides a separate ``GCR 
channel'', which observes penetrating particles with an increased opening angle by omitting the A detectors from the 
coincidence condition, i.e. counting all particles that penetrate B1, C, and B2 (e.g. teal arrow in Fig. \ref{fig:solohet_sensor_schematic}). This leads to an almost 20-fold 
increase in the geometric factor compared to the nominal penetrating particle channel.

For applications requiring even higher counting statistics, it is also possible to use single detector count rates 
without any coincidence conditions, similar to the technique applied e.g. by \citet{Richardson-1996} for the IMP 8 and 
Helios E6 instruments and \citet{Kuehl-2015} for SOHO-EPHIN. In this case, GCR particles are measured from all 
directions (e.g. orange arrow in Fig. \ref{fig:solohet_sensor_schematic}), but without any energy resolution or species separation. The HET C detectors are best suited for this 
purpose due to their large size and nearly isotropic shielding by the aluminum housing. For each HET telescope, 
four such counters are 
available --- each of the two photodiodes has a high-gain channel (C1H, C2H) with a deposited energy 
threshold of $E_\textrm{th}=\SI{4}{\mega\electronvolt}$, and a low-gain channel (C1L, C2L) with 
$E_\textrm{th}=\SI{10}{\mega\electronvolt}$.
As these C detector counters provide no directional information, the values from HET 1 and HET 2 and from the two photodiodes in each telescope can be simply summed up to achieve an even higher count rate, approximately \SI{270}{counts\per\second} for the high-gain channels (C1H + C2H $\times$ 2 units) or \SI{230}{counts\per\second} for the low-gain channels (C1L + C2L $\times$ 2 units). Note that summing up the counts of the two photodiodes does not remove events that were detected in both photodiodes at the same time --- such a counter of all valid events in the C detector is not available in the HET data products and could only be approximated using the Pulse Height Analysis data.

To investigate the response of the HET C counters to an isotropic flux of incoming GCR particles, we have performed a
simulation using Geant4 \citep{Agostinelli-2003}, version 10.1.2, with the physics list QGSP\_BERT. The simulated geometry included a detailed model of the EPT/HET sensor head and the corresponding electronics box, so that the the shielding by the instrument housing and electronics box as well as the generation of secondary particles are taken into account.
A simplified model of the Solar Orbiter spacecraft was also optionally included in the simulation setup to consider the influence of the spacecraft body on the incoming particle flux. This may be important for the C detector counters, as they are sensitive to particles entering HET from any direction. The spacecraft was modeled as a cuboid with the size of the main body ($\SI{2.20}{\meter}\times\SI{1.81}{\meter}\times\SI{1.46}{\meter}$) and total mass of \SI{1700}{\kilogram} (which corresponds to the launch mass of Solar Orbiter, excluding its solar panels). Its composition was assumed to be \SI{200}{\kilogram} of hydrazine fuel, 750 kg of aluminum, representing the structural components of the spacecraft, and 750 kg of a printed circuit board (PCB)-like material, as defined by \citet{Appel-2018} and \citet[Table 6.2]{Appel-2018-PhD}, representing the electronics components of the spacecraft and its payload. The development of a more detailed Geant4 model of the spacecraft body based on CAD models of its components is in progress, but was not possible within the time constraints of this study and is not expected to change the results significantly.
Only protons between \SI{5}{\mega\electronvolt} and \SI{100}{\giga\electronvolt} were used 
as input particles to reduce the complexity of the simulation setup, as protons comprise \SI{90}{\percent} of primary GCR particles \citep{Simpson-1983}.

\begin{figure}
	\centering
	\includegraphics[width=\hsize]{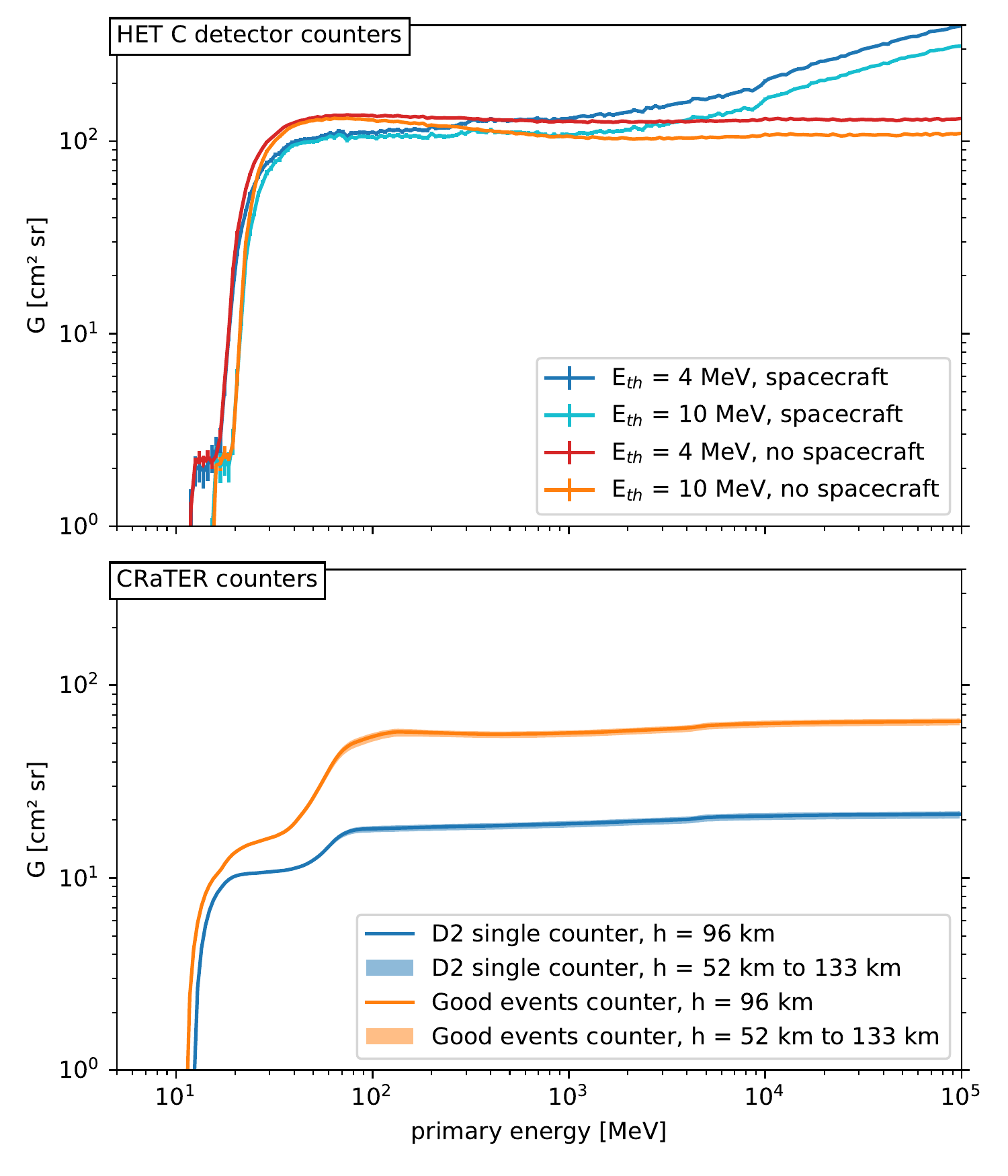}
	\caption{\textit{Upper panel:} Response functions of the SolO HET C detector single counters as determined using a Geant4 simulation. The four lines correspond to four different scenarios depending on the threshold applied for the available counters. The derivation of the response function is described in Sect. \ref{subsec:het}.
	\textit{Lower panel:} Response functions of the D2 detector single counter (blue) and the \textit{good events} counter (orange) of the LRO CRaTER instrument. Lines show the response for the mean altitude of CRaTER during the event, while shaded areas mark the range of responses for the maximum and minimum altitudes. These response functions were derived by \citet{Looper-2013} and are described in Sect. \ref{subsec:crater}.}
	\label{fig:response_functions}
\end{figure}

The proton response function resulting from the simulation is shown in Fig. \ref{fig:response_functions} (upper panel). Four 
curves are shown, corresponding to the simulation setup with and without the spacecraft model, and for the different 
threshold energies of the high- and low-gain channels. It becomes clear that the low-energy cutoff is mainly influenced 
by the threshold energy: \SI{12}{\mega\electronvolt} for the high-gain channel and \SI{16}{\mega\electronvolt} 
for the low-gain channel. After the cutoff follows a narrow plateau corresponding to particles entering C through the nominal field of view (i.e. through the A and B detectors), followed by an increase related to particles entering from the sides through the HET housing.
The spacecraft body provides additional shielding ($\sim\SI{20}{\percent}$) for the detector in the lower energy part, but generates additional secondary 
particles above a primary proton energy of \SI{1}{\giga\electronvolt} --- up to a 2.5-fold increase of the geometric factor for \SI{100}{\giga\electronvolt} particles. On the other hand, without the spacecraft body, 
the geometric factor for high energies stays approximately constant above \SI{1}{\giga\electronvolt}, at
$G=\SI{128+-2}{\centi\meter\squared\steradian}$ for $E_\textrm{th}=\SI{4}{\mega\electronvolt}$ and 
$G=\SI{106+-2}{\centi\meter\squared\steradian}$ for $E_\textrm{th}=\SI{10}{\mega\electronvolt}$. As the GCR proton flux typically peaks at or below \SI{1}{\giga\electronvolt} and decreases again for higher energies, the differences caused by the spacecraft body only have a minor influence on the observed count rates. By folding the response function for $E_\text{th} = \SI{4}{\mega\electronvolt}$ with a typical GCR spectrum at solar minimum ($\Phi = \SI{270}{\mega\volt}$) and integrating over the primary energy, we obtained count rates of \SI{48}{\per\second} without the spacecraft model and \SI{53}{\per\second} with the spacecraft model, an increase on the order of \SI{10}{\percent}. This is only about \SI{80}{\percent} of the typically observed count rate (\SI{270}{\per\second}, divided by 4 channels), as only protons were simulated. We note that the effect of the spacecraft body may be larger for heavier ions, as they fragment more in the spacecraft and may thus contribute more to the response function with the generated secondaries.

\subsection{CRaTER on LRO}
\label{subsec:crater}
The Cosmic Ray Telescope for the Effects of Radiation \citep[CRaTER,][]{Spence-2010} is an instrument on the Lunar 
Reconnaissance Orbiter (LRO) mission measuring the radiation dose and linear energy transfer (LET) spectra in lunar 
orbit. CRaTER consists of three pairs of thin (\SI{140}{\micro\meter}) and thick (\SI{1000}{\micro\meter}) silicon 
detectors, D1 through D6, separated by sections of tissue-equivalent plastic serving as an absorber. The D1 end of the
telescope is pointed towards the zenith, while the D6 end points towards the surface of the Moon. Similarly to HET, 
CRaTER uses multiple coincidence conditions between its detectors to measure particles of different energies. For 
example, the lowest energy particles are detected using the coincidence of D1 and D2 (the uppermost two detectors),
with a minimum energy of \SI{12.7}{\mega\electronvolt} required for protons to penetrate D1 and reach into D2 according 
to \citet{Spence-2010}. This value of \SI{12.7}{\mega\electronvolt} is also the minimum energy for protons to be 
detected in any CRaTER detector, as D1 has a higher energy deposit threshold so that it rejects most protons and many helium ions.

The CRaTER Level 2 secondary science data, available through NASA PDS\footnote{\url{https://pds-ppi.igpp.ucla.edu/}} and on the CRaTER web site\footnote{\url{http://crater-web.sr.unh.edu/}} provide single counters for each of the 6 detectors, similar to 
the HET counters described in Sect. \ref{subsec:het}, as well as additional counters for \textit{rejected events}, 
\textit{good events} and \textit{total events}, where a \textit{good event} is any valid event where an incoming 
particle triggered at least one detector. 

This means that there are two different counters in the CRaTER data (D2 and \textit{good events}) measuring 
protons with energies 
$\gtrsim\SI{12.7}{\mega\electronvolt}$, while the threshold is higher for all other counters. The \textit{good events} 
counter was already used by \citet{Sohn-2019-Forbush,Sohn-2019-Enhancement} to study Forbush decreases and energetic 
particle events, and it has the best counting statistics (on the order of 
$\sim\SI{66}{\per\second}$ at the time of the event studied here). However, while the threshold is well-defined, the response function of the \textit{good events} counter is slightly more complex, as it includes multiple detectors with different shielding conditions and measures a higher amount of secondary particles coming from the lunar surface (the so-called albedo) than D2 alone. \citet[Appendix A]{Looper-2013} have derived the response functions of the single detector count rates using a Geant4 \citep{Agostinelli-2003} simulation. The response functions of the D2 detector single counter as well as the \textit{good events} counter are plotted in \ref{fig:response_functions} (lower panel). Similarly to the HET response function in the upper panel, steps in the response function occur when different parts of the telescope are penetrated by particles.

In addition to the count rate files, we have used the ancillary data of the LRO to exclude time periods where the spacecraft is not in its nominal orientation, e.g. due to orbit adjustment maneuvers. Any data where the LRO is more than \SI{1}{\degree} away from the nominal orientation, with CRaTER's D2 detector pointing towards the zenith, is excluded to make sure that the measured count rates are not affected by these activities. This exclusion only affects few data points, as the LRO pointing is usually very precise to support its imaging instruments.

As the LRO orbit is elliptical and relatively close to the lunar surface (between \SI{54}{\kilo\meter} and \SI{132}{\kilo\meter} above the surface in the time period studied in this work), the Moon 
takes up a significant portion of the sky as viewed from CRaTER. Thus, the Moon shields CRaTER from part of the incoming GCR, but also produces albedo particles. This means that the count rate of 
particles measured using a single-detector counter (i.e. in a $4\pi$ solid angle field of view) periodically varies
with the current altitude, which is also shown in the altitude-dependent response functions in Fig. \ref{fig:response_functions} (lower panel). The plotted altitudes are slightly different from the actual values ($\pm\SI{2}{\kilo\meter}$) due to the limited altitude resolution of the simulation, but this only makes a small difference.
The orbital period of the LRO is about 110 minutes, which determines the frequency of this periodic signal. Multiple methods have been developed to correct for this effect, such as the dose correction factor 
given by \citet{Schwadron-2012} based on geometrical calculation of the covered solid angle, or the Fourier series 
method introduced by \citet{Winslow-2018}. In this study, we apply a simple empirical method in which we create a 
scatter plot of the time-dependent CRaTER count rate $c(t)$ versus the LRO altitude $h(t)$ for the time period of interest, apply a 
linear regression, and use the obtained slope $m$ to calculate the corrected count rate
\begin{equation}
    c_\text{corrected}(t) = c(t) - m \cdot \left(h(t) - \overline{h}\right)
\end{equation}
where $\overline{h}$ denotes the mean altitude of the LRO during the time period investigated, which is \SI{93}{\kilo\meter} for the event studied in this work. In this case, we
have found this method to work about as well as the Fourier series method in suppressing the periodic signal and better
than the simple geometrical calculation, which does not take into account the albedo particles generated by the Moon. 
However, short- or long-term variations of the GCR spectrum, which influence the ratio between the counts of primary GCR 
and albedo particles and thus the necessary correction factor, are not accounted for by any of these methods and can 
still cause the periodic component to appear in the corrected signal, albeit with a much lower amplitude. Due to these
difficulties with the altitude correction, we additionally always plot orbit-averaged values of the CRaTER data.

\subsection{Neutron monitor observations and the global survey method}
\label{subsec:neutronmonitor}

As stated above, neutron monitors have historically been the most important data source for the study of GCR variations in general and FDs in particular. The global network of neutron monitors, whose data are available from the Neutron Monitor Database (NMDB)\footnote{\url{http://www.nmdb.eu}}, provides continuous measurements from many locations around the globe. In contrast to deep space measurements, neutron monitors have an inherent cutoff energy (often given in terms of rigidity) determined by the Earth's magnetosphere and atmosphere, which depends on the latitude as well as the altitude of the neutron monitor. At the poles, the influence of the magnetosphere decreases to zero \cite[see e.g.][]{Smart-Shea-2008}, leading to a cutoff rigidity of \SI{0.1}{\giga\volt} at the location of the South Pole neutron monitor, which would correspond to a proton energy of $\sim\SI{5}{\mega\electronvolt}$. At these locations, the atmospheric cutoff dominates and results in a cutoff energy of about \SI{450}{\mega\electronvolt} for protons \cite{Clem-Dorman-2000}, i.e. a factor of $\sim$ 20--30 larger than in the abovementioned response functions of HET and CRaTER. This causes Forbush decreases observed by neutron monitors to usually be smaller in amplitude than in deep space observations.

A method that takes into account simultaneous ground-level observations of cosmic rays by neutron monitors at different locations to calculate the main characteristics of cosmic-ray variations outside of the atmosphere and magnetosphere of Earth has long been proposed \citep[see][]{krymsky1964diffusion,krymsky1966new,Belov1973,Belov1974,Dorman2009} and is still used nowadays \citep[e.g.][]{Papaioannou2019,Papaioannou2020,Abunina2020}. This technique is called the global survey method (GSM). The GSM separates the isotropic part of the variations of cosmic rays from the anisotropic part and uses spherical harmonics to express their respective amplitudes: In the following, $A0$ is used for the amplitude of the isotropic variations; $Ax$, $Ay$, and $Az$ are the corresponding amplitudes of the first harmonic (higher orders are not considered). $Ax$ and $Ay$ denote the equatorial components of the anisotropy, with $x$ pointing away from the Sun and $y$ perpendicular to that, while $z$ is the north-south component.
However, in order to achieve this, first the atmospheric and instrumental response functions, which couple the primary particles at the top of the atmosphere to the secondaries recorded by neutron monitors on the ground, and a backmapping of cosmic ray particles traveling under the influence of Earth's magnetic field are applied. The historical development, scientific argumentation, and mathematical formulation of GSM can be found in the recent comprehensive report of \cite{Belov2018}.
GSM incorporates a power-law dependence on the rigidity for the isotropic part of the CR variations (i.e. $A0$) and thus can provide outputs for a set of fixed rigidities \citep[see for example Figure 2 in][]{2000SSRv...93...79B}. However, a fixed rigidity of \SI{10}{\giga\volt} (corresponding to a proton energy of \SI{9.1}{\giga\electronvolt}) has typically been used for more than 65 years \citep[e.g.][]{2000SSRv...93...79B, Belov2015, Belov2018, Papaioannou2020, Abunina2020}. This value is more illustrative on the actual GCR modulation and is close to the effective rigidity of NMs to detect GCRs \citep[see e.g.][]{2017JGRA..122.9790A,2018SoPh..293..110K}, implying that a NM is mostly responsive to the variability of mid-rigidity CRs from several GV to several tens of GV in rigidity.
 
\section{Methods}
\label{sec:models}

\subsection{ForbMod}
\label{subsec:forbmod}

ForbMod \citep{Dumbovic-2018-ForbMod} is an analytical physics-based model to describe Forbush decreases caused by flux rope CMEs. Its calculations are based on the self-similar expansion of a flux rope, which is modeled as a (locally) cylindrical structure with an initial radius $a_0$ close to the Sun 
that initially contains no GCRs at its center. While the flux rope propagates away from the Sun, it expands self-similarly: Both the increase of the flux rope radius $a$ and the decrease of the central magnetic 
field magnitude $B_\text{c}$ are assumed to follow power law expressions with the so-called expansion factors $n_a$ and 
$n_B$ used as power law indices:
\begin{equation}
    a(t) = a_0 \left(\frac{R(t)}{R_0}\right)^{n_a}, \qquad B_\text{c}(t) = B_0 \left(\frac{R(t)}{R_0}\right)^{-n_B},
    \label{eq:forbmod_powerlaw}
\end{equation}
where $R(t)$ describes the radial distance of the flux rope from the Sun, $R_0$ the initial distance at time $t=0$, and $B_0$ the initial central magnetic field. 
As stated by \citet{Dumbovic-2018-ForbMod}, previous observational studies 
\citep{Bothmer-Schwenn-1998,Leitner-2007,Demoulin-2008,Gulisano-2012} constrained the power law indices to 
$0.45 < n_a < 1.14$ and $0.88 < n_B < 1.89$.
During the expansion and radial propagation of the CME, GCRs gradually diffuse into the flux rope slower than in the surrounding solar wind, so that the GCR phase space density within the flux rope is decreased while it passes by an observer. 
ForbMod then describes the GCR phase space density within the flux rope using the following main equations, which are 
derived in detail by \citet{Dumbovic-2018-ForbMod}:
\begin{equation}
    U(r, t) = U_0 \left(1 - J_0\mathopen{}\left(\alpha_1 \frac{r(t)}{a(t)}\right)\mathclose{} \; e^{-\alpha_1^2 f(t)} \right),
    \quad
    f(t) = \frac{D_0}{a_0^2} \left(\frac{v}{R_0}\right)^x \frac{t^{x+1}}{x+1}
    \label{eq:forbmod}
\end{equation}
where $U_0$ is the GCR phase space density outside the flux rope, $J_0$ is the Bessel function of first kind and order zero, $\alpha_1$ is a constant corresponding to the first positive root of $J_0$, $r$ is the radial distance of the observer from the flux rope center (which may be time-dependent, hence $r(t)$), $D_0$ is the initial diffusion coefficient and $v$ is the CME propagation speed. The function $f(t)$ describes the GCR diffusion into the flux rope, where the diffusion time is equivalent to the propagation time $t$ since the initial condition ($t=0$) near the Sun. It is assumed that $v$ is constant and that the diffusion coefficient $D$ is inversely proportional to the central magnetic field, $D \propto 1/B_\text{c}$, so that $D(t)$ follows a power law with the index $n_B$ (c.f. Eq. \ref{eq:forbmod_powerlaw}). This power law relation was already inserted to obtain the expression for $f(t)$ given in Equation \ref{eq:forbmod}.
Additionally, the ambient GCR phase space density $U_0$ is assumed to be constant to simplify the calculation; the known radial gradient the GCR flux of about \SI{3}{\percent\per\AU} \citep{Webber-1999,Gieseler-2016,Lawrence-2016} is not taken into account. The expansion type
\begin{equation}
    x = n_B - 2n_a \neq -1
    \label{eq:expansion_type}
\end{equation}
describes the expansion behavior of the CME, and in particular its magnetic flux. $x = 0$ corresponds to a conservation of magnetic flux (as the product of the flux rope cross section and the central magnetic field stays constant), while $x > 0$ describes a decrease of the flux with heliospheric distance and $x < 0$ an increasing flux. $x = -1$ is a special case, which requires a different functional form of $f(t)$ in Equation \ref{eq:forbmod} \citep[for details, see][]{Dumbovic-2018-ForbMod}. The influence of the value of $x$ on the ForbMod result can be understood as the interplay between the expansion and diffusion effects -- when the diffusion (which depends on the magnetic field, and thus, $n_B$) is very efficient, the flux rope is quickly filled with GCR particles, but a fast increase of the flux rope size (large $n_a$) can counteract this effect by increasing the space that needs to be filled with GCRs.

In addition to its dependence on the magnetic field, the GCR diffusion coefficient $D$ also depends on the 
particle energy. E.g., higher energy particles diffuse into the flux rope more easily and thus show a shallower FD. 
While the original model of \citet{Dumbovic-2018-ForbMod} describes only the FD profile of one specific GCR energy, 
for which $D_0$ needs to be provided, \citet{Dumbovic-2020-ForbMod} extended the model with empirical functions for the 
energy dependence of the diffusion coefficient, so that the FD profile can be calculated for any GCR energy. By folding 
the resulting spectrum with the response function of a particle detector, it is then possible to simulate the measurement of the FD by this detector.
In this version of ForbMod, the input GCR spectrum and the energy dependence of the diffusion coefficient $D$ are needed as input parameters for the model. As described by \citet[Appendix B]{Dumbovic-2020-ForbMod}, the modified 
force-field approximation described by \citet{Gieseler-2017} is used to calculate the GCR spectrum based on the values 
of the solar modulation potential $\Phi$ obtained from neutron monitor data by \citet{Usoskin-2011} and from ACE/CRIS 
data by \citet{Gieseler-2017}. For our event in April 2020, near the minimum between Solar Cycles 24 and 25, the
corresponding measurements of $\Phi$ are not yet available, so we use the values from similar conditions for the previous 
solar cycle in June 2009. The values from \citet{Usoskin-2011} are derived based on data from the Oulu neutron monitor --- as its count rates between April 2020 and June 2009 are comparable, this supports our assumption that the solar modulation conditions are very similar. The energy-dependent diffusion coefficient is calculated using the empirical formula given by 
\citet{Potgieter-2013}, with parameters derived by \citet{Potgieter-2014} for the period 2006--2009 from PAMELA data 
and by \citet{Corti-2019} for the period 2011--2017 from AMS-02 measurements. In this case, data for 2020 are not yet 
available either, so we use the values from 2009 with comparable solar cycle conditions. More detailed explanations 
about these parameters are given by \citet[Appendix A]{Dumbovic-2020-ForbMod}.

To convert the $U(r, t)$ dependence in Equation \ref{eq:forbmod} into a function that purely depends on the time $t$ and thus can directly be compared to in situ GCR measurements, the observer location $r$ with respect to the flux rope center needs to be defined. For this, we can use the in situ measured velocity profile $v_\text{in situ}(t)$ of the flux rope, i.e. the observer passes through the flux rope at this measured velocity:
\begin{equation}
r(t) = \big| a(t) - v_\text{in situ}(t) \cdot (t - t_\text{CME}) \big|
\label{eq:forbmod_timeprofile}
\end{equation}
where $t_\text{CME}$ is the in situ arrival time of the CME. The conversion of $U(r)$ into $U(t)$ introduces some asymmetry into the FD profile, as the in situ measured velocity profile $v_\text{in situ}(t)$ is typically not constant.

Note that ForbMod only models the GCR modulation due to a flux rope CME, not the additional influence of a shock/sheath region, although it may be combined with other models to take this into account \citep[see e.g.][]{Dumbovic-2020-ForbMod,Forstner-2020}.

\section{Results}
\label{sec:results}

\subsection{In situ observations}
\label{subsec:insitu_observations}

\begin{figure*}
    \sidecaption
    \includegraphics[width=12cm]{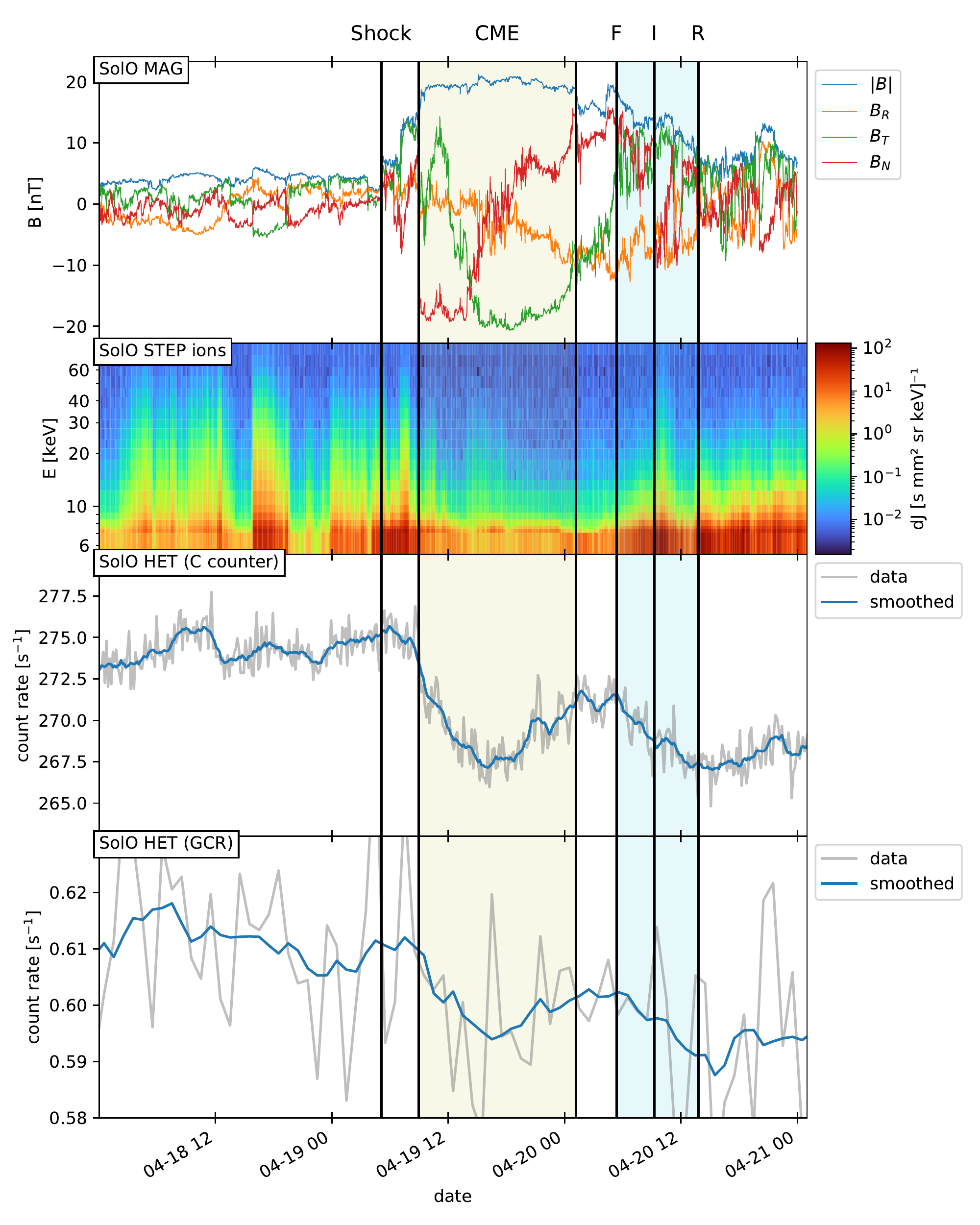}
    \caption{Measurements from MAG, STEP and HET on Solar Orbiter, showing the magnetic structure of the CME, 
        suprathermal particle signatures, and the associated FD observations in the GCR channel and the C detector counter of HET.
        The MAG measurements are displayed in radial (R), tangential (T) and normal (N) coordinates.
        Black vertical lines and shaded regions mark the time periods corresponding to different events: Shock arrival, CME (flux rope) start and end, as well as forward shock (F), stream interface (I) and reverse shock (R) of the SIR.}
    \label{fig:cme_data_at_solo}
\end{figure*}

The April 19 CME was observed at Solar Orbiter using its magnetometer, showing a clear signature 
of a flux rope with a south-east-north field rotation and a maximum field intensity of $B_\textrm{max}=\SI{21.2}{\nano\tesla}$, a 
preceding shock with a jump of about \SI{3}{\nano\tesla} in magnetic field intensity, and a turbulent sheath region
in between (see the upper panel in Fig. \ref{fig:cme_data_at_solo}, and see \citet{Davies-2021} for further discussion of the MAG data). The shock arrival time was 05:06 UTC on April 19, 
2020, the flux rope arrived at 08:58 UTC on the same day and ended at 01:11 UTC on April 20. 
MAG data from April 21 (i.e. one day after the end of the CME flux rope) are not displayed here because spacecraft commissioning activities affected the sensor temperatures on that day.
Solar wind plasma measurements from the Solar Wind Analyzer instrument on SolO \citep[SWA,][]{Owen-2020-SWA} are not available for this event, as it was not yet fully commissioned.
EPD measured the fluxes of suprathermal ions slightly above solar wind energies 
(\SIrange{5.3}{85}{\kilo\electronvolt}, i.e. \SIrange{1000}{4000}{\kilo\meter\per\second}) using the SupraThermal 
Electrons and Protons (STEP) instrument. As shown in the second panel of Fig. \ref{fig:cme_data_at_solo}, STEP sees a clear enhancement of suprathermal ions 
accelerated in the sheath region, and this is also confirmed by EPD's Electron Proton Telescope (EPT, not shown here), which saw enhancements of ions up to \SI{100}{\kilo\electronvolt}. No significant enhancements of energetic electrons were observed in EPT or STEP.

\begin{figure*}
	\sidecaption
	\includegraphics[width=12cm]{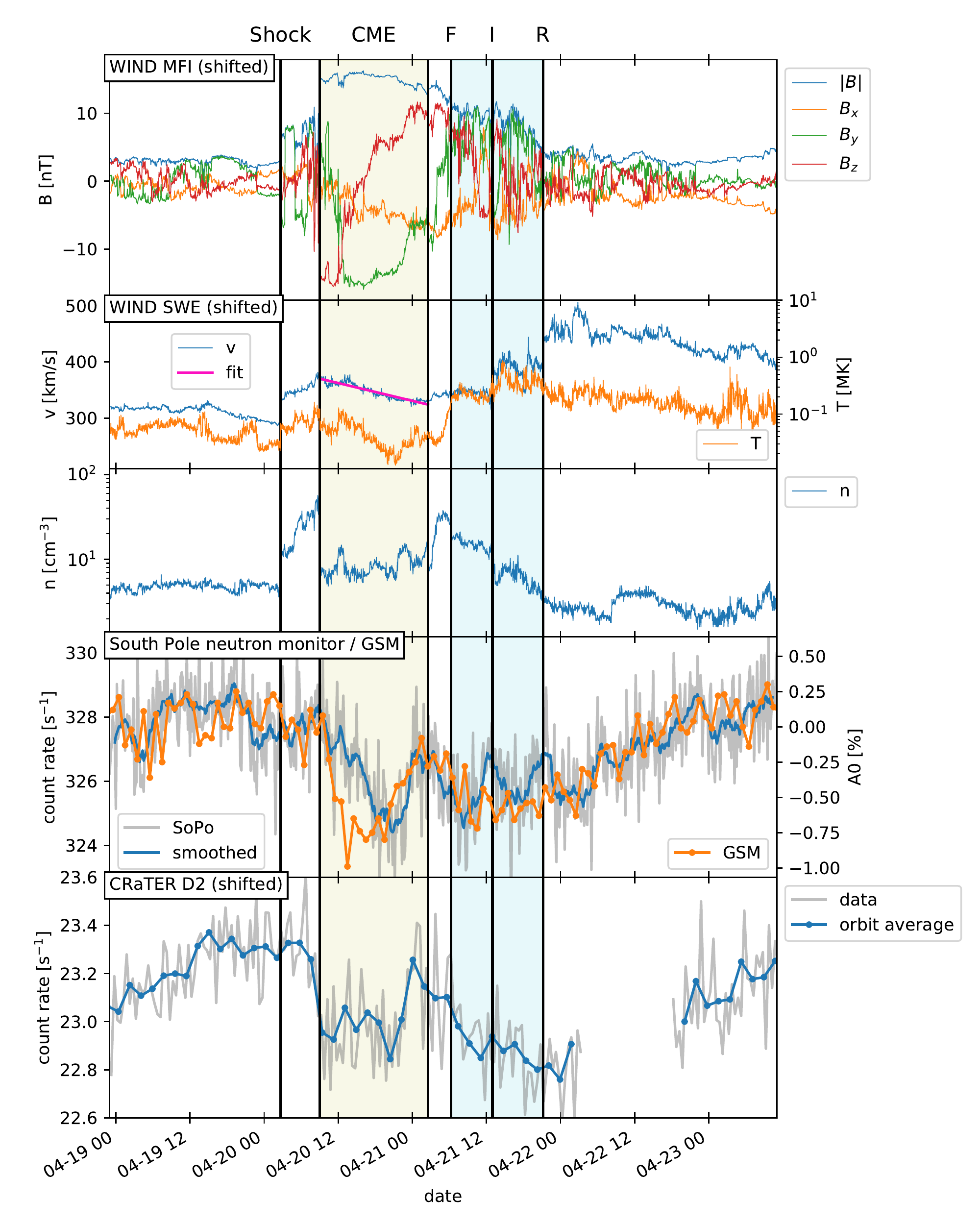}
	\caption{Measurements near Earth from MFI and SWE on Wind and neutron monitors on Earth as well as the CRaTER D2 counter, showing the in situ signatures of the CME and the associated FD, as well as a high speed stream following afterwards. Wind data were shifted forward in time by 1 hour and 7 minutes to account for the expected transit time between the L1 Lagrange point and Earth, and CRaTER data were shifted by 15 minutes, corresponding to the Moon--Earth radial distance.
		Black vertical lines and shaded regions mark the time periods corresponding to different events: Shock arrival, CME start and end, as well as forward shock (F), stream interface (I) and reverse shock (R) of the SIR.
     		Wind MFI measurements are given in Heliocentric Earth Ecliptic (HEE) coordinates, with X pointing from the Sun to Earth and Z being perpendicular to the ecliptic pointing north, and Y completing the right handed triad. The general orientation of HEE is thus comparable to RTN, which is used for SolO data in Fig. \ref{fig:cme_data_at_solo}, and the difference to RTN is small.
		A linear fit to the velocity profile of the flux rope, which is used to determine the expansion speed as    explained by \citet{Gulisano-2012}, is shown in pink.
		The second to bottom panel shows both measurements from the south pole neutron monitor (gray, blue) and the GCR density variation at \SI{10}{\giga\volt} (corresponding to \SI{9.1}{\giga\electronvolt} proton energy) obtained from GSM.}
	\label{fig:cme_data_at_1au}
\end{figure*}

The flux rope is followed by a separate structure with enhanced levels of magnetic turbulence. By comparing with the solar wind plasma observations near Earth (see Fig. \ref{fig:cme_data_at_1au} and its description later in this section), where clear increases in solar wind 
speed and temperature are observed, we have identified this to be a stream interaction region (SIR), followed by a stream of high speed solar wind. We have determined the onset times of the three SIR structures, the forward shock (F), stream interface (I), and reverse shock (R) at SolO by searching for shock signatures in the magnetic field data that are similar to those seen at Wind, though the identification is less reliable than at Earth due to the missing SWA data. STEP and EPT also see another enhancement of energetic ions close to the stream interface.

However, the main focus of this study is the signature in the high energy particles, where a clear Forbush decrease
with a drop amplitude of around \SI{3}{\percent} in both the GCR channel as well as the C detector counters is observed (bottom panels of Fig. \ref{fig:cme_data_at_solo}). The C counter is plotted in 10 minute time averages, with an additional curve showing a smoothed version of these data (rolling mean), and the GCR channel is shown in a similar fashion with a 1 hour cadence.
Due to the higher
count rate, the FD is especially well observed in the C counters. The main part of the decrease occurs during the passage
of the flux rope --- the decrease within the sheath region is well below \SI{1}{\percent}. This means that the assumption of the ForbMod model (Sect. \ref{subsec:forbmod}) that only the flux rope effect is taken into account is fulfilled. A second GCR decrease is observed after the CME, coinciding with the passage of the SIR.

Fig. \ref{fig:cme_data_at_1au} shows the in situ measurements of the CME arrival near Earth, including solar wind 
magnetic field and plasma data from the Magnetic Field Investigation \citep[MFI,][]{Lepping-1995-WindMFI} and the Solar 
Wind Experiment \citep[SWE,][]{Ogilvie-1995-WindSWE} onboard the Wind spacecraft, as well as GCR measurements from the 
South Pole neutron monitor (SoPo), the GSM outputs and the CRaTER D2 counter.
The measured speed of the CME at Wind is very 
slow at a maximum of $\SI{370}{\kilo\meter\per\second}$.
The Wind and CRaTER measurements were shifted forward in time taking into account the radial distance to Earth (\SI{1}{\hour} \SI{7}{\minute} for Wind at L1, and \SI{15}{\minute} for CRaTER at the Moon --- cf. the inset in Fig. \ref{fig:solar_system}). These time shifts were calculated using the abovementioned maximum speed of $\SI{370}{\kilo\meter\per\second}$, which is seen at the front of the CME flux rope.
Considering this time shift, the shock arrival time at Earth is 02:40 UTC on April 20, and the flux rope arrived a few hours later at 09:01 UTC. In comparison to Solar Orbiter, the magnetic field strength of the flux rope has decreased to a maximum of $B_\textrm{max}=\SI{16.2}{\nano\tesla}$, while the sheath region still has field intensities similar to the SolO measurement around \SI{6}{\nano\tesla}.
Apart from the lower intensity, the magnetic field signatures of the flux rope look very similar to those observed at Solar Orbiter, showing the same south-east-north orientation. The large negative out-of-ecliptic component ($B_Z$, or $B_N$) seen at the beginning of the flux rope is a feature which is typically associated with high geoeffectiveness \citep[e.g.][]{Gopalswamy-2008}.
The sheath duration increased by more than two hours ($\sim\SI{60}{\percent}$), which is probably related to the accumulation of additional solar wind plasma in front of the CME as well as expansion due to the increasing velocity profile of the sheath region \citep[see e.g.][]{Manchester2005,Siscoe2008,Janvier-2019,Forstner-2020}, while the expansion of the flux rope is more moderate at a bit more than one hour ($\sim\SI{8}{\percent}$).
The transit times from Solar Orbiter to L1 correspond to an average transit speed of \SI{363}{\kilo\meter\per\second} for the flux rope front, which matches the in situ measured front speed of \SI{370}{\kilo\meter\per\second} very well.

As mentioned before, the SIR following the CME is clearly seen in the in situ data at Wind, showing signatures such as the increases in temperature and velocity as well as a decrease in density. According to these signatures, the time of the forward shock (F), stream interface (I) and reverse shock (R) were marked in Fig. \ref{fig:cme_data_at_1au}. Even though the solar wind plasma data are not available at Solar Orbiter for this event (as described above), it is clear from the magnetic field measurements that the SIR followed closely behind the CME at both locations, separated by a region of high plasma density (seen in the Wind measurements).
Assuming an average solar wind speed within the SIR of approximately \SI{400}{\kilo\meter\per\second}, the separation of the Parker spiral footpoints of SolO and Earth is \SI{16.7}{\degree}, corresponding to an expected SIR delay time of 27.2 hours. The measured delay varies between 24.9 hours for the forward shock, 27.7 hours for the stream interface, and 31.4 hours for the reverse shock, suggesting that the SIR has significantly expanded in both directions.
This means that the SIR may have affected the evolution of the CME, e.g. by compressing it from behind. We will discuss this further in Sect. \ref{sec:discussion_conclusions}.

Comparing the GCR measurements at SoPo and CRaTER, as well as the outputs of GSM, when utilizing measurements of $\sim$35 neutron monitors, it can be seen that the relative amplitudes of the FD profiles induced by the CME at SoPo and from GSM are quite similar, whereas both are quite different compared to CRaTER. As discussed in Sect. \ref{subsec:crater}, CRaTER covers a similar energy range as the HET C counter at SolO, while neutron monitors have a larger cutoff energy. The South Pole neutron monitor has much higher counting statistics than CRaTER, but the FD there only reaches an amplitude of \SI{1.2}{\percent}, as higher energy particles are modulated less by the CME's magnetic field. This is also true for the outputs of GSM that reach an amplitude of \SI{1.1}{\percent}. The minimum of the FD appears to fall within the magnetic cloud (MC), and as at SolO, the MC seems to be the main driver of the FD in comparison to the shock/sheath structure, during which only a small decrease is observed. On the other hand, the FD at CRaTER has an amplitude of \SI{2.0}{\percent}. The FD onset at CRaTER appears to be slightly earlier than the arrival of the flux rope, but the difference is only less than one orbital period of CRaTER, so this may also be an artifact of the altitude correction (cf. Sect. \ref{subsec:crater}). The slightly enhanced periodic variations of the CRaTER signal seen close to the minimum of the FD are also a sign that the altitude correction is not completely suppressing the periodic signal due to the modulated GCR spectrum.

\begin{figure}
	\includegraphics[width=\hsize]{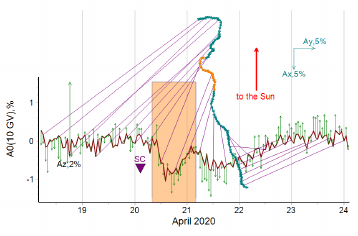}
	\caption{GCR density variation $A0$ at Earth obtained from GSM at a fixed rigidity of \SI{10}{\giga\volt} (brown line, corresponding to \SI{9.1}{\giga\electronvolt} proton energy), together with the first harmonic of the cosmic ray anisotropy. The equatorial component $Axy$ of the anisotropy is displayed as a vector diagram (teal and orange triangles), which are connected to the corresponding points in time on the $A0$ plot with magenta lines. Additionally, the north-south component $Az$ is shown as green vertical arrows on top of $A0$ time profile. The shaded rectangle and the orange part of the vector diagram correspond to the duration of the magnetic cloud (MC). The shock arrival at Earth is indicated by the arrow labeled SC (``sudden storm commencement"), and the direction to the Sun in the vector diagram is indicated with a red arrow. The components of the anisotropy $Ax$ and $Ay$ that define the plane for the calculation of $Axy$ are indicated on the top right corner of the figure. Numbers at each anisotropy component on the figure indicate the scale used for the plotting of the relevant arrows.}
	\label{fig:anisotropy}
\end{figure}

Fig. \ref{fig:anisotropy} presents the density variations of cosmic rays at Earth obtained from GSM: $A0$ (in \%), together with the components of the anisotropy $Axy$ (equatorial components) and $Az$ (polar component). The characteristics of the cosmic ray anisotropy that signify the effect of a MC on GCRs are summarized as follows: \textit{(a)} the amplitude of $Axy$ is higher within the MC, reaching a maximum of $\sim\SI{1}{\percent}$ coinciding with the minimum of the FD; \textit{(b)} the direction of the anisotropy vector (i.e. orange part of the vector diagram) changes abruptly when entering the MC \citep{Belov2015}; \textit{(c)} there is a rotation of the $Axy$ vector within the MC, and \textit{(d)} the north-south component $Az$ changes by \SI{1.1}{\percent} during the FD, including a reversal of direction during the decay phase of the FD \citep{Abunin2013,Belov2015}.


The CME parameters calculated from the in situ data at Solar Orbiter and near Earth as well as the onset times of the SIR structures are summarized in Table \ref{tab:insitu_data}.

\begin{table}
	\caption{CME and SIR parameters derived from the in situ measurements at Solar Orbiter and near Earth.}
	\label{tab:insitu_data}
	\centering
	\def\arraystretch{1.2}
	\begin{tabular}{rrr}
		\hline\hline
		& Solar Orbiter & near Earth   \\ \hline
		\multicolumn{3}{l}{Radial distance}                  				 \\
		\hline
		$R$ [AU]					     & 0.809				  & 0.995\tablefootmark{a} / 1.005\tablefootmark{b} \\
		\hline
		\multicolumn{3}{l}{CME and SIR onset times}                                      \\
		\hline
		$t_\text{shock}$ [UTC]           & 2020-04-19 05:06       & 2020-04-20 02:40 \\
		$t_\text{CME}$ [UTC]             & 2020-04-19 08:58       & 2020-04-20 09:01 \\
		$t_\text{CME end}$ [UTC]         & 2020-04-20 01:11       & 2020-04-21 02:32 \\
		$t_\text{forward shock}$ [UTC]   & 2020-04-20 05:22\tablefootmark{c} & 2020-04-21 06:15 \\
		$t_\text{stream interface}$ [UTC] & 2020-04-20 09:15\tablefootmark{c} & 2020-04-21 12:58 \\
		$t_\text{reverse shock}$ [UTC]   & 2020-04-20 13:47\tablefootmark{c} & 2020-04-21 21:11 \\
		\hline
		\multicolumn{3}{l}{Duration}						                 \\
		\hline
		$\Delta t_\text{sheath}$ [h]            & 3.9                    & 6.4              \\
		$\Delta t_\text{CME}$ [h]               & 16.2                   & 17.5             \\
		\hline
		\multicolumn{3}{l}{In situ parameters}			                     \\
		\hline
		$B_\text{max}$ [nT]              & 21.2                   & 16.2             \\
		$\overline{v}_\text{CME}$ [km/s] & ---                    & 347              \\
		$v_\text{exp}$ [km/s]            & ---                    & 46               \\
		$A_\text{FD}$ [\%]               & 2.9                    & 2.0              \\ \hline
	\end{tabular}
	\tablefoot{The listed Forbush decrease amplitudes $A_\text{FD}$ correspond to the HET C counter at Solar Orbiter and the CRaTER D2 counter near Earth.}
	\tablefoottext{a}{L1}
	\tablefoottext{b}{Earth}
	\tablefoottext{c}{Due to the missing plasma data, SIR onset times are less certain at SolO.}
\end{table}

\begin{figure*}
	\includegraphics[width=\hsize]{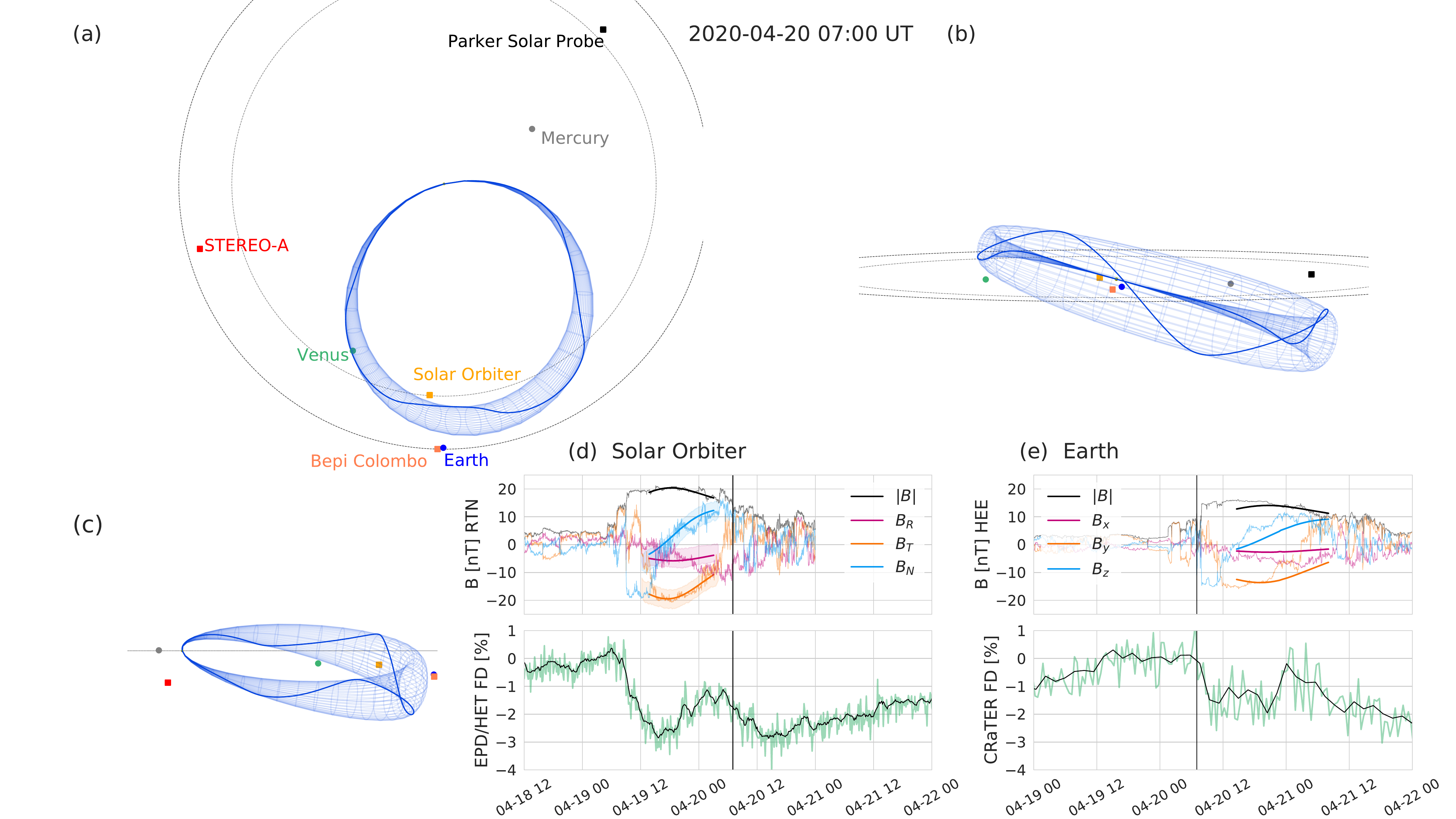}
	\caption{Visualization of the results of the 3DCORE flux rope model fitted to the Solar Orbiter MAG observations, shown at the time of the onset of the Forbush decrease at Earth. The reconstructed 3D flux rope structure is shown (a) looking down from the solar north pole onto the solar equatorial plane, (b) in a frontal view along the Sun--Earth line, and (c) in a side view at a 75 degree angle, the longitude of STEREO--A to Earth. A flux rope field line is highlighted as a solid blue line. The panels (d) and (e) show the in situ magnetic field data from Solar Orbiter and Wind at Earth/L1 compared to the GCR variation as a percentage drop in the amplitude measured by EPD/HET and CRaTER. The Wind magnetic field components are given here in Heliocentric Earth Ecliptic (HEE) coordinates, as in Fig. \ref{fig:cme_data_at_1au}. The 3DCORE modeled magnetic field is overplotted in panel (d) and propagated to Earth as shown in (e). An animation of this figure is available as an \textbf{online movie}. 
	}
	\label{fig:3dcore}
\end{figure*}

Fig. \ref{fig:3dcore} (also available as an online animation) shows the results from an application of the semi-empirical 3DCORE model \citep{Moestl2018-3DCORE, Weiss2020-3DCORE}, based on the SolO MAG observations of the flux rope. This model provides a global context for the flux rope structure, propagation and orientation at Solar Orbiter and Wind. Further applications of this model and its results are described in more detail by \citet{Davies-2021}, and here we show a few main results relevant for our study. In order to reconstruct the magnetic field configuration and its 3D structure, the 3DCORE model ensembles were fitted to an interval of the MAG data with a clean magnetic field rotation, and shown is a run that represents a best fit which covers an interval from Apr 19 11:13 UTC to Apr 20 01:59 UTC. This interval starts about 2 hours later than the start of the flux rope interval stated above at 08:58 UT as the $B_T$ component is positive for a short while after 09:00 UT, which is inconsistent with its unipolar excursion to $B_T<0$ later. This first feature in $B_T$ cannot be fitted with the 3DCORE flux rope model, and thus we choose to narrow the fitting interval to what \citet{Davies-2021} call the ''unperturbed'' inner part of the flux rope.

The 3DCORE technique consists of a Gold-Hoyle uniform twist magnetic field in an elliptical flux rope cross section placed in a 3D toroidal shape \citep{Weiss2020-3DCORE}. Here, we set the cross section aspect ratio, otherwise a free parameter to be determined from the fitting analysis, to a value of 2.0, which is consistent with the angular width of the CME void in Heliospheric Imager observations \citep{Davies-2021}. In Fig. \ref{fig:3dcore}a-c, a 3D visualization of the 3DCORE envelope is presented from several viewpoints at the time of the Forbush decrease onset at Earth. Fig.~\ref{fig:3dcore}d demonstrates the ability of the model to fit the Solar Orbiter observations. In Fig. \ref{fig:3dcore}e, we propagated the model to the Wind spacecraft self-similarly, where the power law exponents for the expansion of the diameter and magnetic field (as defined in Equation \ref{eq:forbmod_powerlaw}) were set to previously empirically derived values of $n_a=1.14$ and $n_B=1.64$, respectively. Those values are based on a power law fit of the mean total magnetic field of a large sample of in situ measured CME flux ropes in the inner heliosphere \citep{Leitner-2007}. 

The 3DCORE torus propagates according to a drag-based model \citep[see details in][]{Weiss2020-3DCORE}. The results show that the modeled magnetic field components are consistent at Solar Orbiter and Wind at L1, but as seen in Fig. \ref{fig:3dcore}e, there is a time shift between the model and the observations of the flux rope magnetic field at Wind (concerning all components and the total field). This points to a slight inconsistency of the Solar Orbiter fit results when they are propagated to L1, which most likely arises from the shape and direction of the 3DCORE torus being determined with data from a single spacecraft, and it is expected that due to the model assumptions this does not exactly reproduce the observations at another spacecraft. This inconsistency can be alleviated with simultaneously fitting 3DCORE to Solar Orbiter and Wind in situ magnetic field data, but this is the subject of future studies. 

In Fig. \ref{fig:3dcore} we show a model which uses parameters representative of the best fit, but with the fitting algorithm we use \citep{Weiss2020-3DCORE} we can derive distributions for each of the flux rope parameters. The main results from the Solar Orbiter 3DCORE fit are, with the results stated as means $\pm$ standard deviations: the CME is directed at \SI{13 \pm 5}{\degree} longitude (HEEQ) and \SI{-5 \pm 5}{\degree} latitude, which means that it has a close to central impact at Solar Orbiter and Wind and the observations at the two spacecraft are clearly connected.  The orientation of the axis is \SI{11 \pm 13}{\degree} to the solar equatorial plane, thus it is a low inclination flux rope. At the heliocentric distance of Wind (\SI{0.995}{\AU}), the axial magnetic field strength in the model is \SI{14.3 \pm 0.9}{\nano\tesla}, and the model flux rope has a diameter of \SI{0.114 \pm 0.022}{\AU}. For Solar Orbiter at \SI{0.809}{\AU} this axial field is \SI{20.1\pm 1.2}{\nano\tesla} and the diameter is \SI{0.090 \pm 0.017}{\AU}.


As explained by \citet{Demoulin-Dasso-2009} and \citet{Gulisano-2012}, the measured velocity profile of the flux rope can be used to estimate the expansion factor $n_a$, which describes the increase of the flux rope radius $a$ with the radial distance from the Sun (see definition in Sect. \ref{subsec:forbmod}). From a linear fit, we calculate the expansion speed $v_\text{exp}$, which is the velocity difference between the front and rear end of the flux rope, to be \SI{46}{\kilo\meter\per\second}, and together with the mean speed of $\overline{v}_\text{CME} = \SI{347}{\kilo\meter\per\second}$ we calculate:
\begin{equation}
    n_{a, \textrm{in situ @ Wind}} = \frac{v_\text{exp}}{\Delta t_\text{CME}} \frac{R}{\overline{v}_\text{CME}^2} = 0.90
    \label{eq:n_a_local}
\end{equation}
where $R$ is the radial distance of Wind at this time (see \ref{tab:insitu_data}).

It is also possible to calculate a value of $n_a$ for the propagation between SolO and Wind, using the two in situ measurements:
\begin{equation}
	n_{a, \textrm{SolO-Wind}} = \log \left(\frac{\Delta t_\text{CME, Wind}}{\Delta t_\text{CME, SolO}}\right) \bigg/ \log \left(\frac{R_\text{Wind}}{R_\text{SolO}}\right) = 0.37 \,,
	\label{eq:n_a_solowind}
\end{equation}
and, similarly, we can derive the expansion factor $n_B$ for the magnetic field magnitude between SolO and Wind:
\begin{equation}
n_{B, \textrm{SolO-Wind}} = -\log \left(\frac{B_\text{max, Wind}}{B_\text{max, SolO}}\right) \bigg/ \log \left(\frac{R_\text{Wind}}{R_\text{SolO}}\right) = 1.30 \,,
\label{eq:n_B_solowind}
\end{equation}
The values for both $n_{B, \textrm{SolO-Wind}}$ and $n_{a, \textrm{in situ @ Wind}}$ are within the typical ranges found in previous observational studies, as described in Sect. \ref{subsec:forbmod}. The value $n_{a, \textrm{SolO-Wind}}$ is unusually low and quite different from the in situ measurement. This could be interpreted as a sudden change in the expansion rate of the CME, but may also be related to the difference of the inherent assumptions in the two methods: For example, the local determination of the expansion factor at Wind (Eq. \ref{eq:n_a_local}) assumes a quasi-undisturbed expansion of the CME following the current velocity profile within the flux rope, while external influences are not taken into account. Contrarily, the observation of a SIR that follows closely behind the CME, as described above, suggest that there may have been some interaction between the two structures that may have affected the expansion. This will be discussed in more detail in Sec. \ref{subsec:forbmod_application} and \ref{sec:discussion_conclusions}.
We also note that the derived values of $n_a$ and $n_B$ are both lower than the fixed values assumed in the 3DCORE model, but this is partly due to the fact that the 3DCORE modeling excludes the first part of the flux rope duration (as explained above). Also, as stated above, a more detailed 3DCORE analysis fitting the CME structure simultaneously at both locations will be explored in future studies.

\subsection{Remote sensing observations}
\label{subsec:remotesensing_observations}

\begin{figure*}
	\centering
	\includegraphics[width=\hsize]{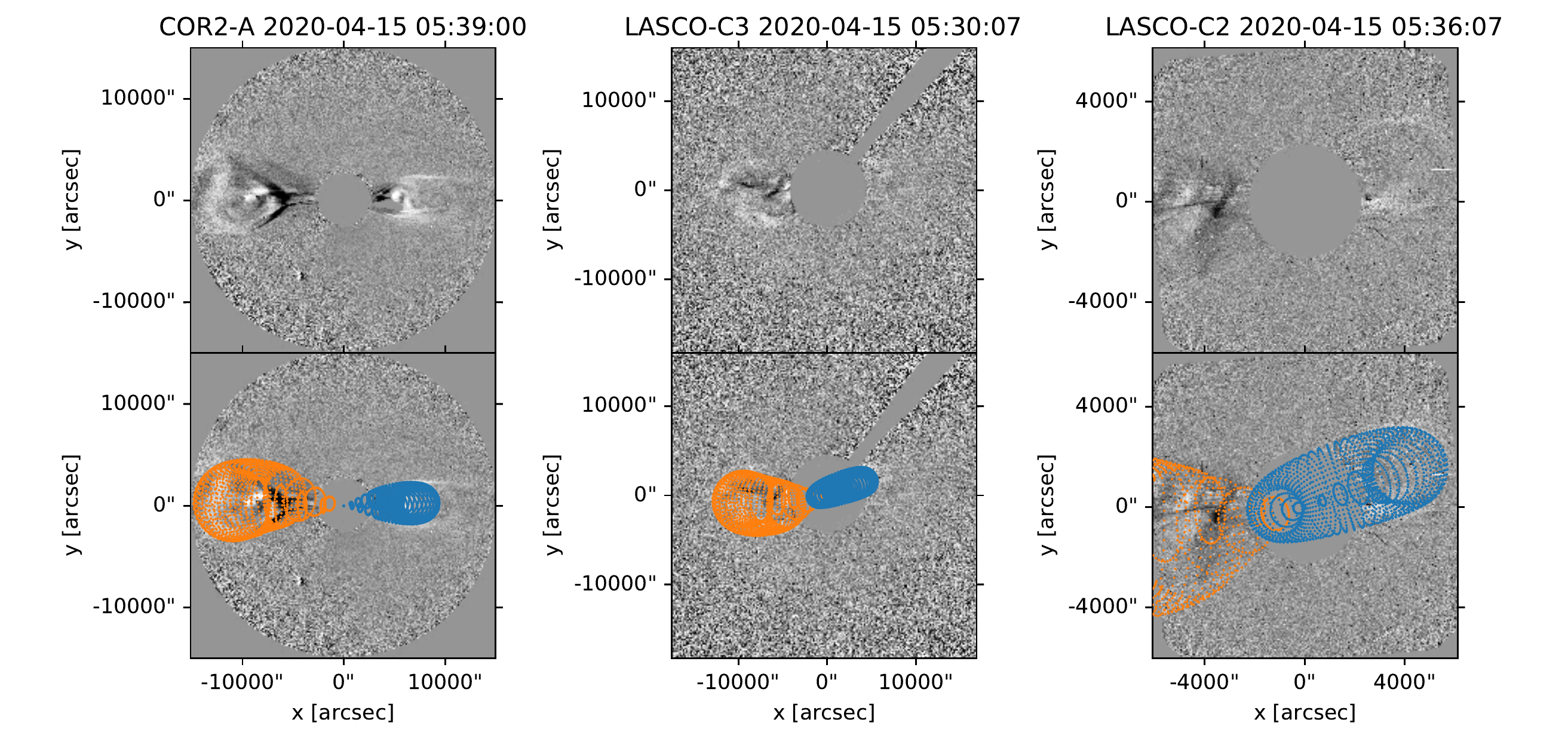}
	\includegraphics[height=0.35\hsize]{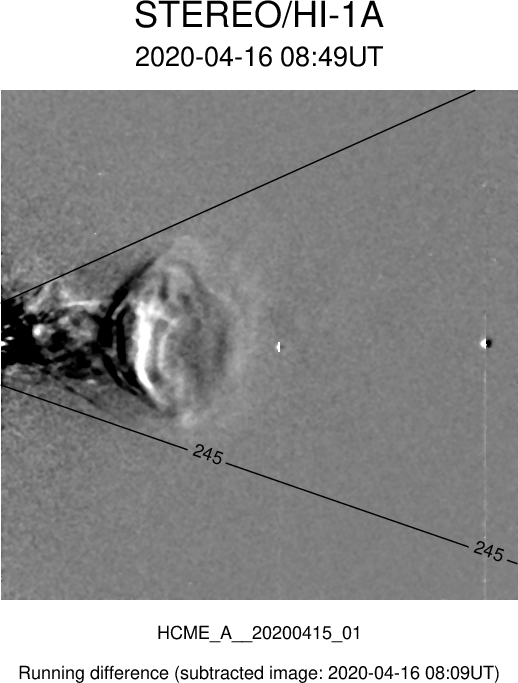}
	\includegraphics[height=0.35\hsize]{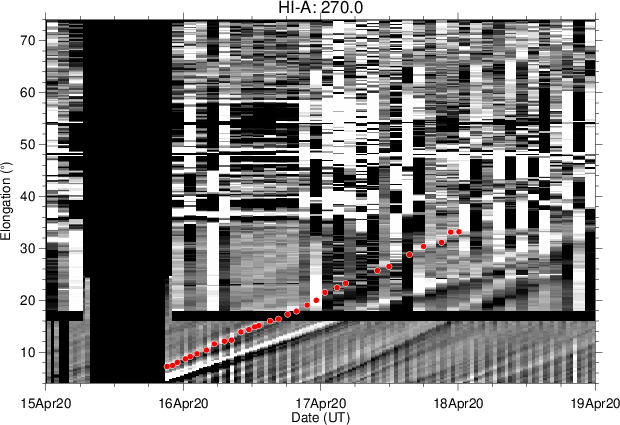}
	\caption{Remote sensing observations of the CME. In the STEREO-A COR2 and SOHO LASCO C2 and C3 running difference images (top part), the GCS fitting was applied to derive the parameters $R_0$ and $a_0$ for the ForbMod model (see results in Table \ref{tab:gcs_fit}). The blue markings denote the Earth-directed CME we are investigating, while the CME fitted in orange is backsided and launched a few hours earlier. STEREO-A HI observations (bottom left) and time-elongation maps (bottom right) are provided by the HELCATS project.}
	\label{fig:remotesensing_images}
\end{figure*}

Due to its \SI{75}{\degree} longitudinal separation from Solar Orbiter and Earth at the time (cf. Fig. \ref{fig:solar_system}), the STEREO-A spacecraft 
has provided excellent remote-sensing observations of this CME event. Fig. \ref{fig:remotesensing_images} shows 
observations from the Sun Earth 
Connection Coronal and Heliospheric Investigation suite onboard STEREO-A \citep[SECCHI,][]{Howard-2008-SECCHI}, namely the 
COR2 white-light coronagraph, as well as from the Heliospheric Imagers (HI).
The COR2 image shows two CMEs launched from the Sun in close succession on April 14--15 2020. The CME visible on the right side of the COR2 image, which first appears at approximately 19:30 UTC on April 14 and then slowly moves outward, is the one that headed towards SolO and Earth, while the larger CME on the left side is backsided from the Earth point of view.

To reconstruct the CME shape near the Sun, we have applied the Graduated Cylindrical Shell model \citep[GCS,][]{Thernisien-2006-GCS,Thernisien-2011-GCS} to the STEREO-A/COR2 and SOHO/LASCO C2 and C3 \citep{Brueckner-1995-LASCO} coronagraph images, which allows us to derive parameters such as latitude and longitude as well as the flux rope height and radius. In the process of this study, we have developed a new implementation of the GCS model in Python\footnote{\url{https://github.com/johan12345/gcs_python},\\ \url{https://doi.org/10.5281/zenodo.4443203}}, and verified its results against the existing SolarSoft IDL version. During the reconstruction process, it became apparent that the structure seen on the east limb from SOHO/LASCO C3 cannot belong to the Earth-directed CME. To fit the GCS geometry to this structure, it would have been necessary to shift the CME longitude by more than \SI{30}{\degree} away from Earth and/or increase the flux rope height significantly, which contradicts the position of the clear flux rope structure observed at STEREO-A/COR2 and the in situ observation at Earth and Solar Orbiter.
Considering this, we suspect that this signature is instead caused by the backsided CME, and we have verified this by also approximately fitting the backsided CME with the GCS model (as plotted in orange in Fig. \ref{fig:remotesensing_images}).
The Earth-directed CME is not clearly seen in the LASCO C3 images, but shows a weak signature in C2 on the northwestern limb. This structure was used in conjunction with the clear observations in the STEREO-A COR2 data to reconstruct the CME (plotted in blue in Fig. \ref{fig:remotesensing_images}). The GCS results show that the two CMEs partly overlap in the SOHO/LASCO observations due to the line of sight effect, which is probably the reason why the Earth-directed CME is only seen from SOHO on the west limb. The fit parameters for both CMEs are listed in Table \ref{tab:gcs_fit}, where the uncertainties were derived by performing the GCS fit for the Earth-directed CME 40 times and then calculating the mean and standard deviation of each parameter. This was not done for the backsided CME as its parameters are not needed for the further analysis in this study.
The GCS fit results for the latitude, longitude and tilt angle are also approximately consistent with the data derived from the 3DCORE reconstruction based on the in situ data (see Sect. \ref{subsec:insitu_observations}, though these are of course also associated with some uncertainties.
We note that the 40 GCS fits of the Earth-directed CME were performed by a single person, which may decrease the uncertainties compared to a result produced using independent reconstructions from different scientists. Still, care was taken to sample a large range of possible values for each parameter and adjust the remaining parameters accordingly to fit the coronagraph images. Additionally, the data were compared to a single independent GCS reconstruction by another researcher, and the results agree within the given uncertainty ranges.

There is also no obvious signature of the CME in the low corona (low coronal signatures, LCS), as observed with the SDO/AIA \citep{Lemen-2012-AIA} \SI{211}{\angstrom} extreme ultraviolet (EUV) images, making it challenging to identify the CME source region. A weak brightening is observed at approximately \SI{2}{\degree}N \SI{8}{\degree}E, but this is too far away from the GCS-reconstructed CME longitude of \SI{18+-7}{\degree}W. Thus, the CME can be considered as a type of ``stealth CME'' \citep[see e.g.][and references therein]{Howard-2013-StealthCMEs} both due to the weak LCS and the lack of a clear halo CME in the coronagraphs from the Earth point of view. Stealth CMEs have weak LCS because only a relatively small amount of energy is released from the corona at their onset, due to their low speed (typically $<\SI{300}{\kilo\meter\per\second}$ according to \citet{Ma-2010}), and these signatures may be too weak to be detected with the established observational and data processing techniques \citep[e.g.][]{Alzate_2017}. A more detailed study of the source region of this CME will be performed by O'Kane et al. (2021, in preparation for \textit{A\&A}).

As a result of the GCS fit, we derived the initial height of the flux rope $R_0 = \SI{9.64+-0.4}{\solarradius}$ and the initial radius at the apex $a_0 = \SI{1.93+-0.15}{\solarradius}$, calculated using the equation from \citet{Thernisien-2011-GCS}. These parameters will be needed for the application of the ForbMod model in Sect. \ref{subsec:forbmod_application}.

\begin{table}
	\caption{Results from the graduated cylindrical shell (GCS) model.}
	\label{tab:gcs_fit}
	\centering
	\def\arraystretch{1.2}
	\begin{tabular}{rS[table-format=2.2(2)]S[table-format=3.2]}
		\hline\hline
		                                 & {CME 1\tablefootmark{a}} & {CME 2\tablefootmark{b}}  \\ \hline
		HEEQ Longitude [\si{\degree}]    & 18+-7                    & 229                       \\
		HEEQ Latitude [\si{\degree}]     & 3+-3                     & -1                        \\
		Tilt angle [\si{\degree}]        & 18+-6                    & 6                         \\
		Half angle [\si{\degree}]        & 35+-11                   & 15                        \\
		Height [$R_\odot$]               & 9.5+-0.8                 & 14.73                     \\
		Ratio $\kappa$                   & 0.23+-0.04               & 0.30                      \\ \hline
		FR radius at apex [$R_\odot$]    & 1.8+-0.3                 & 3.78                      \\ \hline
	\end{tabular}
	\tablefoot{GCS fitting was applied in the 2020-04-15 05:39:00 UTC image from STEREO-A COR2 and the 2020-04-15 05:36:07 UTC image from SOHO/LASCO C2. Results are plotted in Fig. \ref{fig:remotesensing_images}. Error bars are given only for CME 1, as CME 2 is not further studied here.}
	\tablefoottext{a}{Directed towards SolO and Earth}
	\tablefoottext{b}{Backsided as seen from SolO/Earth}
\end{table}

Based on the GCS results, we can make a new calculation for the expansion factor $n_a$: The calculation in Equation \ref{eq:n_a_local} corresponds to the instantaneous expansion of the flux rope near \SI{1}{\AU}, which may not be the same as closer to the Sun. The average expansion factor between the Sun and Earth can be calculated by comparing the initial flux rope size $a_0$ with the one measured in situ at Wind:
\begin{equation}
n_{a, \textrm{Sun-Wind}} = \log \left(\frac{a_\text{Wind}}{a_0}\right) \bigg/ \log \left(\frac{R_\text{Wind}}{R_0}\right) = 0.70
\label{eq:n_a_global}
\end{equation}
where $R_\text{Wind}$ is the radial distance of the Wind spacecraft from the Sun and $a_\text{Wind} = \Delta t_\text{CME} \cdot \overline{v}_\text{CME} / 2 = \SI{15.7}{\solarradius}$ is the flux rope radius calculated from the in situ data (see Table \ref{tab:insitu_data}). A similar value of $n_a = 0.69$ can be calculated from the SolO measurements, when assuming the CME speed to be the same as at Wind.

STEREO-A HI observations clearly show the CME signature out to elongation angles of approximately 
\SI{35}{\degree} (corresponding to a radial distance of $\sim$\SI{0.6}{\AU}), as seen in the running difference images and the time-elongation map (Fig. \ref{fig:remotesensing_images}, bottom panels). This event is cataloged by the HELCATS project\footnote{\url{https://www.helcats-fp7.eu/}} under the ID \texttt{HCME\_A\_\_20200415\_01}.
According to the self-similar expansion fitting (SSEF) result \citep{Davies-2012-SSE} given in the HELCATS HIGeoCat catalog \citep{Barnes-2019}, the CME direction in Heliocentric Earth Equatorial (HEEQ) coordinates is \SI{-6}{\degree} in longitude and \SI{-2}{\degree} in latitude. The longitude does not match what we determined in our GCS reconstruction (Table \ref{tab:gcs_fit}) --- but as the SSEF technique only uses data from a single spacecraft and makes certain assumptions about the CME, such as a constant speed and a fixed half-width of \SI{30}{\degree}, it is known to often produce large uncertainties for the CME longitude \citep[see e.g.][]{Barnes-2019}.
The SSEF results can also be used to calculate the arrival time at Solar Orbiter and Earth, as described by \citet{Moestl-2017-HelcatsHSO}.
The calculated arrival times available from the ARRCAT v2.0\footnote{\url{https://helioforecast.space/arrcat},\\ \url{https://doi.org/10.6084/m9.figshare.12271292}} are 2020-04-19 09:10 $\pm$ 3.2 h for SolO and 2020-04-20 09:45 $\pm$ 4.0 h for L1, which are both about 8 hours later than the in situ shock arrival times. This is well within a usual range of arrival time errors with this method of $\pm 17$ hours. \citep{Moestl-2017-HelcatsHSO}.
The arrival speed at Earth is predicted as \SI{335+-11}{\kilo\meter\per\second}, which is also consistent with the in situ measured CME speed (mean speed $\overline{v} = \SI{347}{\kilo\meter\per\second}$, see Table \ref{tab:insitu_data}).

\subsection{Application of the ForbMod model}
\label{subsec:forbmod_application}

In the previous sections \ref{subsec:insitu_observations} and \ref{subsec:remotesensing_observations}, we have derived all input parameters necessary for applying the ForbMod model to the Forbush decreases at Solar Orbiter and Earth. For the flux rope radius expansion factor $n_a$, multiple values were calculated from measurements at different locations, with quite significant differences: $n_{a, \textrm{SolO-Wind}} = 0.37, n_{a, \textrm{in situ @ Wind}} = 0.90$ and $n_{a, \textrm{Sun-Wind}} = 0.70$. Additionally, we have derived one value of the magnetic field expansion factor $n_{B, \textrm{SolO-Wind}} = 1.30$, based on the SolO and Wind in situ measurements of the magnetic field.

While $n_B$ is also used separately to derive the radial dependence of the diffusion coefficient $D$, the key purpose of the two expansion factors within ForbMod, which makes the model very sensitive to these values, is to calculate the so-called expansion type, a quantity defined as $x = n_B - 2n_a$
(see equation \ref{eq:expansion_type}). It describes the evolution of the magnetic flux and is assumed to be constant over the course of the CME propagation, i.e. the magnetic flux increases or decreases at the same rate.
Thus, the inconsistency of the measured $n_a$ values suggests that $n_B$ must also have changed to keep $x$ constant. So, we can derive $x = n_{B, \textrm{SolO-Wind}} - 2 n_{a, \textrm{SolO-Wind}} = 0.55$, and then, under the assumption that $x = \text{const.}$, calculate a corresponding $n_B$ for each of the measured $n_a$ values. The results of this calculation are listed in Table \ref{tab:forbmod_expansion}.
Of course, in the case of this event, $x = \textrm{const.}$ is a quite bold assumption to make considering the observed variation of $n_a$, but due to the lack of additional observations of $n_B$, there is no other way to derive the necessary input parameters from observations. We will discuss the possible implications of this in more detail in Sect. \ref{sec:discussion_conclusions}.

\begin{table*}
    \caption{Input parameters for the ForbMod model}
    \label{tab:forbmod_parameters}
    \centering
	\def\arraystretch{1.5}
	\begin{tabular}{p{4cm}p{5cm}cp{4cm}}
		\hline\hline
		Parameter &
		Source &
		Section &
		Value \\ \hline
		GCR spectrum &
		Force-field approximation, \citet{Gieseler-2017} &
		\ref{subsec:forbmod} &
		$\Phi$ for June 2009 \\
		Diffusion coefficient $D$ &
		Empirical function from \citet{Potgieter-2013} with parameters from \citet{Potgieter-2014} &
		\ref{subsec:forbmod} &
		parameters for 2009 \\
		Detector response function &
	 	Geant4 simulation results &
		\ref{subsec:het}, \ref{subsec:crater} &
		See Fig. \ref{fig:response_functions}\\
		Magnetic field $B_\text{c}$ &
		$B_\maxt$ in Wind data &
		\ref{subsec:insitu_observations} &
		$B_\text{c} = \SI{16.2}{\nano\tesla}$ \\
		expansion factors $n_a, n_B$ &
		Calculation assuming $x = \text{const.}$ (Eq. \ref{eq:expansion_type}) / best fit &
		\ref{subsec:forbmod_application} &
		see Table \ref{tab:forbmod_expansion} \\
		Flux rope parameters $R_0, a_0$ &
		GCS reconstruction &
		\ref{subsec:remotesensing_observations} &
		$R_0 = \SI{9.64}{\solarradius}$,\newline $a_0 = \SI{1.93}{\solarradius}$ \\
		Diffusion time \newline ($\approx$ transit time) &
		In situ arrival time, Launch time: time of GCS fit&
		\ref{subsec:insitu_observations}, \ref{subsec:remotesensing_observations} &
		$t_\text{SolO} = \SI{99}{\hour}$,\newline $t_\text{Earth} = \SI{123}{\hour}$ \\
		Velocity profile &
		linear fit to in situ measurements at Wind &
		\ref{subsec:insitu_observations} &
		see Table \ref{tab:insitu_data} \\
		\hline
	\end{tabular}
\end{table*}

\begin{table}
	\caption{Pairs of expansion factors $n_a, n_B$ used for the ForbMod model, and resulting FD amplitudes at SolO HET, CRaTER and the South Pole neutron monitor.}
	\label{tab:forbmod_expansion}
	\centering
	\def\arraystretch{1.5}
	\begin{tabular}{rp{1.2cm}p{1.2cm}p{1.2cm}p{0.8cm}}
		\hline
		Calculation                & Sun $\to$ Wind        & SolO $\to$ Wind & in situ @ Wind        & best fit\tablefootmark{a} \\
		\hline
		$n_a$                      & 0.70                  & 0.37            & 0.90                  & 1.08     \\
		$n_B$                      & 1.95\tablefootmark{b} & 1.30            & 2.36\tablefootmark{b} & 2.01     \\
		$x$ \tablefootmark{c}      & \textit{0.55}         & 0.55            & \textit{0.55}         & -0.15     \\
		\hline
		$A_\text{FD, SolO}$ [\%]   & < 0.01                & < 0.01          & 1.25                  & 2.90     \\
		$A_\text{FD, CRaTER}$ [\%] & < 0.01                & < 0.01          & 0.50                  & 2.00     \\
		$A_\text{FD, SoPo}$ [\%]   & < 0.01                & < 0.01          & 0.03                  & 0.44     \\
		\hline
	\end{tabular}
	\tablefoot{Each column in the table corresponds to one set of input parameters $n_a$ and $n_B$ that was used with ForbMod. The modeled FD amplitude for the GSM data (\SI{10}{\giga\volt}) is $<\SI{0.01}{\percent}$ for all four sets of input parameters and not shown here.}
	\tablefoottext{a}{Best fit was obtained by constraining the FD amplitudes at SolO and CRaTER.}
	\tablefoottext{b}{These quantities were calculated assuming that $x = 0.55$ (see discussion in Sect. \ref{subsec:forbmod_application}).}
	\tablefoottext{c}{Calculated using Equation \ref{eq:expansion_type}.}
\end{table}

To summarize, we list all the parameters that are used for the application of ForbMod again in Table \ref{tab:forbmod_parameters}. We have run ForbMod for each of the $n_a$ and $n_B$ pairs that we calculated (Table \ref{tab:forbmod_expansion}), as well as for a ``best fit'' result reproducing the measured FD amplitudes at Solar Orbiter HET and CRaTER. Apart from the response functions, transit times and radial distances, the ForbMod input parameters were always the same for both locations. It also needs to be noted that following the observed variation of $n_a$, the duration of the FD profile calculated with ForbMod was not derived from the $a(t)$ power law assumed by ForbMod (equation \ref{eq:forbmod_powerlaw}), but instead was fixed to the observed flux rope duration.
The ForbMod best fit was obtained by calculating the FD amplitudes across the whole reasonable parameter space of $n_a$ and $n_B$ (while keeping all other parameters fixed) and then selecting the set of parameters that produced the lowest sum of squared residuals with respect to the two in situ measured amplitudes at HET and CRaTER (see Table \ref{tab:insitu_data}).

\begin{figure*}
    \sidecaption
    \includegraphics[width=0.7\hsize]{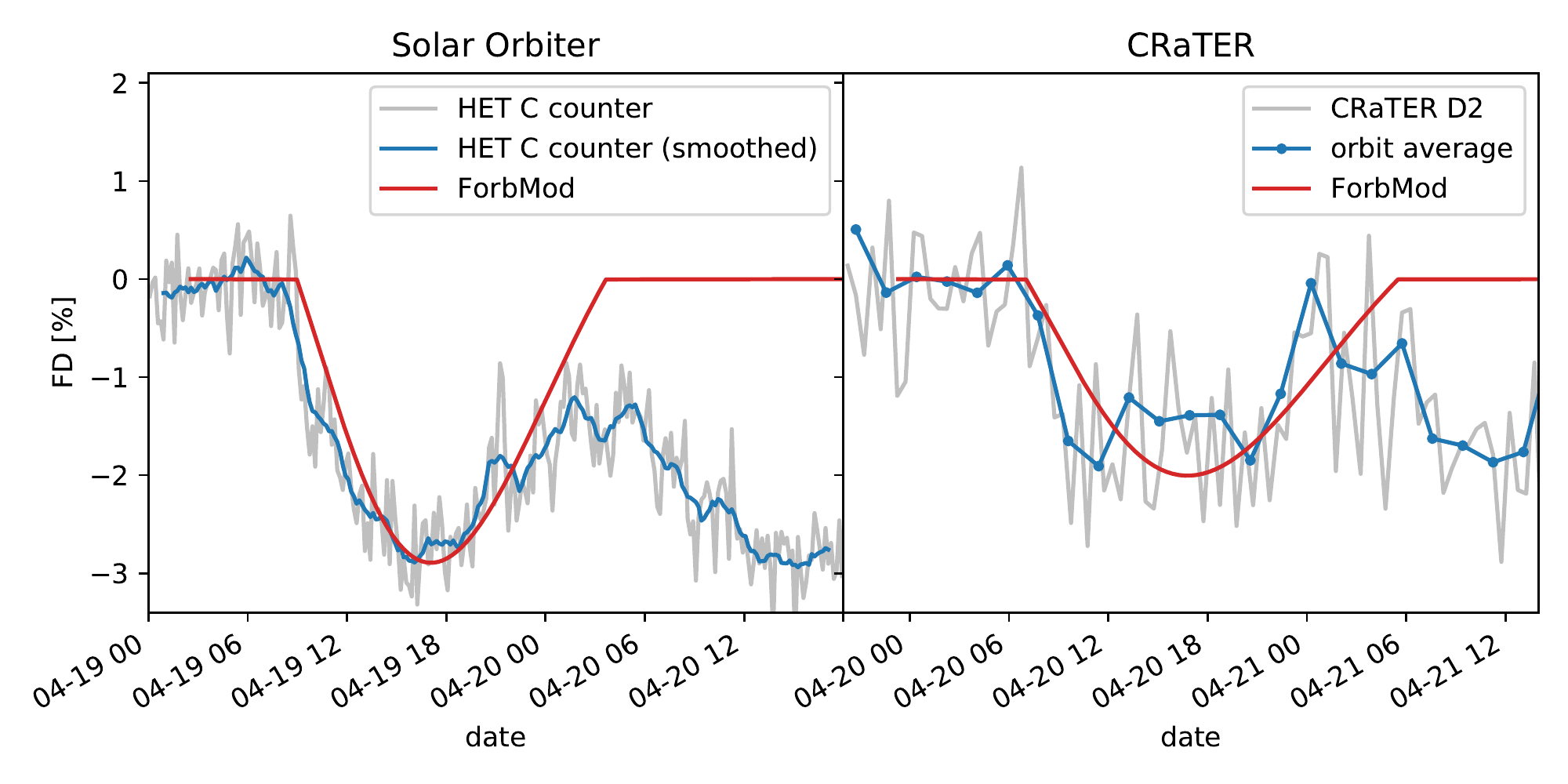}
    \caption{ForbMod model results at Solar Orbiter and CRaTER, in comparison with the measured Forbush decreases at Solar Orbiter and CRaTER. The measurements are plotted in the same fashion as in Figures \ref{fig:cme_data_at_solo} and \ref{fig:cme_data_at_1au}, but normalized to their pre-onset values to reflect the relative variation of the GCR count rate. Input parameters for ForbMod are listed in Tables \ref{tab:forbmod_parameters} and \ref{tab:forbmod_expansion}. $n_a$ and $n_B$ were used from the ``best fit'' result.}
    \label{fig:forbmod}
\end{figure*}

The ForbMod results for the ``best fit'' parameters are shown in Fig. \ref{fig:forbmod}, where the time profile calculated using Equation \ref{eq:forbmod_timeprofile} is plotted in red and the measurements in blue/gray (as previously in Figures \ref{fig:cme_data_at_solo} and \ref{fig:cme_data_at_1au}). It can be seen that for these parameters, there is a good agreement between the model and observations: ForbMod describes well the relatively symmetric Forbush decrease caused by the flux rope CME and reproduces the observed FD amplitudes. Of course, the second decrease caused by the SIR is not included in the model, which explains the obvious deviation of the measurements from the model after the flux rope passage. The effect of the spacecraft model included in the HET response function (see Sect. \ref{subsec:het}) is significant, applying ForbMod using the response function without the spacecraft would lead to a $\sim\SI{20}{\percent}$ larger FD (amplitude of \SI{3.52}{\percent}, not shown here).

For the other parameters $n_a$ and $n_B$ derived from the observations, ForbMod results for the FD amplitude at SolO and CRaTER are shown in Table \ref{tab:forbmod_expansion}. With all these parameter sets, it can be seen that ForbMod underestimates the amplitude of the FD. The closest result is obtained using the in situ parameters measured at Wind, but even in this case the modeled FD amplitude is less than half of the measurement. For the other sets of parameters, ForbMod predicts the flux rope to already be completely filled with GCRs by the time it reaches SolO and Earth, so that the FD amplitude is $<\SI{0.01}{\percent}$.

In addition to SolO HET and CRaTER, we have applied ForbMod at Earth with different response functions to model the FDs observed at the South Pole neutron monitor (SoPo) and in the GSM data. For the latter, we have applied ForbMod monoenergetically at the fixed rigidity of \SI{10}{\giga\volt} (corresponding to \SI{9.1}{\giga\electronvolt} proton energy) that is used by GSM, while for the former we assume a constant response above a cutoff energy of \SI{450}{\mega\electronvolt} (see Sect. \ref{subsec:neutronmonitor}). With these results, both the FDs at SoPo and GSM data are significantly underestimated, with a maximum amplitude of \SI{0.44}{\percent} for SoPo and well below \SI{0.01}{\percent} for GSM in all cases. To obtain the observed FD amplitude on the order of \SI{1}{\percent} from the model, especially for the higher energy of GSM, the parameters $n_a$ and/or $n_B$ would need to be increased even more, which is not supported by observations or the previous observational constraints cited in Sect. \ref{subsec:forbmod}.

\section{Discussion and conclusions}
\label{sec:discussion_conclusions}

In this study, we have shown in situ and remote sensing observations of the first flux rope CME that hit Solar Orbiter on April 19 and Earth on April 20, 2020, and studied the Forbush decrease that it caused at both SolO and near Earth.
This event is considered to be a ``stealth CME'' as it showed only weak signatures from the Earth point of view in the EUV observations of the low corona. Remote sensing observations of this CME were only possible thanks to the ideal position of the STEREO-A spacecraft, which could track the CME from the outer corona until $\sim$\SI{0.6}{\AU}.
At SolO and Earth, the CME was followed by a SIR, which is also clearly observed at both locations in the in situ magnetic field and cosmic ray signatures.

The largest FDs in terms of magnitude often require the presence and combined effect of the shock-sheath region and the following ejecta, both of which are necessary for deep GCR depressions \citep[e.g.][]{Papaioannou2020}. Additionally, CMEs that are characterized as magnetic clouds (MCs) are more often associated with large FDs \citep{2011SoPh..270..609R}.  However, although the CME studied in this work was associated with a shock and a clear MC observed at both SolO and Earth, the resulting amplitude of the FD at Earth was relatively low, especially at higher GCR energies measured by neutron monitors and the GSM (only $\sim$\SI{1}{\percent}). This may lead to the belief that only the flank of the CME hit SolO and Earth, but this is not supported by the in situ flux rope modeling in Sect. \ref{subsec:insitu_observations} and the remote sensing observations presented in Sect. \ref{subsec:remotesensing_observations}. Instead, the slow propagation speed of the CME below \SI{350}{\kilo\meter\per\second} led to a very weak shock and an extremely long propagation time of more than five days from the Sun to Earth, which made it possible for GCRs to diffuse into the MC and thus decrease the observed FD amplitude. This explanation relates well to the concept of the ForbMod model, because the diffusion of GCRs into the flux rope over time is the basis for its calculations. Based on the timing of the FD, is seems that the shock only had a very weak effect on the GCR modulation, with the main part being caused by the MC.

Using input parameters derived from the CME observations, we have applied the ForbMod model to reproduce the FDs observed by the SolO HET C detector counter and the LRO CRaTER D2 counter. Most parameters were relatively straightforward to derive. Only the expansion factors $n_a$ and $n_B$, which describe the evolution of the flux rope radius and its magnetic field and to which the model is quite sensitive, could not be unambiguously determined from observations, as they seem to vary significantly depending on where and how they are measured. In addition, while $n_a$ could be measured for different locations, $n_B$ could only be measured based on SolO and Wind observations (i.e. there is no measurement of $n_B$ from the Sun to Earth). A set of parameters derived assuming that the expansion type $x$ is constant and using the $n_a$ value measured in situ near Earth only produces a FD amplitude of $\SI{1.29}{\percent}$ at SolO compared to the measurement of $\sim\SI{3}{\percent}$. A ``best fit'' set of parameters $n_a$ and $n_B$, which closely reproduce the FD amplitudes measured at CRaTER and HET, was also calculated, with an even higher value of $n_a$, i.e. a stronger expansion of the flux rope size not supported by the observations, and a lower value of $x$ closer to zero, corresponding to a conserved magnetic flux \citep[see][]{Dumbovic-2018-ForbMod}. In this case, the faster expansion (larger $n_a$) counteracts the diffusion of GCRs into the flux rope, as described in Sect. \ref{subsec:forbmod}, so that the FD amplitude becomes larger. However, even with the ``best fit'' parameter values, the higher-energy FD measurements of the South Pole neutron monitor and the global survey method (GSM) could not be reproduced with ForbMod.

Using the observation of the flux rope evolution from SolO to Earth in ForbMod yields FDs which do not agree with the observations. In addition, ForbMod best fit parameters yield global flux rope Sun-to-Earth evolutionary parameters which are far away from the values derived based on in situ measurement comparison between SolO and Earth. This might indicate that Sun-to-Earth evolution of this CME was different from the SolO-to-Earth evolution. A similar event, a slow stealth CME followed by a high speed stream was studied by \citet{He-2018}. They have shown that the CME was compressed by the fast solar wind behind it, which caused an enhanced magnetic field and thus an unexpectedly high geoeffectiveness. The same may have happened for this event --- the inconsistent measured values of $n_a$, which correspond to a slower expansion of the flux rope between the Sun and \SI{1}{\AU} than suggested by the in situ measured velocity profile, can be a result of such a compression, and thus indicate that the expansion behavior of this very slow CME may have changed during its propagation time. E.g. at some point during its propagation, the CME may have been slightly compressed by the SIR, and expanded more freely at other times. Consequently, the assumption of ForbMod that the flux rope radius and its magnetic field follow power laws with constant indices $n_a$ and $n_B$ and that the resulting expansion type $x$, which describes the evolution of the magnetic flux, is also constant may not be valid in this more complex case. This may well be the reason why the model is not able to reproduce the higher energy FD measurements, even with a set of input parameters that fits the lower energy measurements of HET and CRaTER.

Another possible explanation for this discrepancy is that the energy dependence of the ForbMod-modeled FD amplitude may simply be overestimated for this event, resulting in too low FD amplitudes at higher energies. This could happen e.g. if the empirical input parameters for the GCR spectrum and the energy dependence of the diffusion coefficient do not match the actual conditions at this time. This is an interesting result and should be investigated in more detail in future studies. E.g., a statistical validation of ForbMod against the results of the GSM, which has already been applied to a large catalog of FDs, may be helpful to examine whether this is a systematic problem in the description of the energy dependence for these higher GCR energies, or whether this disagreement is a specific attribute of this CME event due to its low speed, very long propagation time, and possible influence of the following SIR.

This study highlights the capabilities of the instruments onboard the Solar Orbiter spacecraft, such as the high counting statistics of the HET C detector capable of detecting Forbush decreases. In addition, it shows that coordinated observations with Solar Orbiter and other spacecraft will be extremely important for the better understanding of space weather in the inner heliosphere. Spacecraft close to the Sun, such as Solar Orbiter and Parker Solar Probe, can serve as an upstream monitor to provide valuable information and early warning about CMEs.
The CME in this case study also serves as an excellent example for a ``stealth CME'' that was still geoeffective due to its strong magnetic field even though it was not clearly seen in remote sensing observations from the Earth point of view. This again highlights that the monitoring of Earth-directed CMEs requires in situ and remote sensing measurements at additional locations, such as at Solar Orbiter and Parker Solar Probe as well as from STEREO-A or a future L5 mission.
As Solar Orbiter's trajectory moves closer to the Sun in the coming years and the solar activity increases with the commencement of Solar Cycle 25, space weather events during conjunctions with Earth as well as other spacecraft will become more probable, which will provide more exciting science opportunities.

\begin{acknowledgements}
J. v. F. thanks L. Seimetz and N. Lundt for their assistance in simulating the HET detector response functions. Additionally, we thank M. D. Looper and J. Wilson from the CRaTER team for providing the response functions of their instrument and helpful suggestions about the analysis of the CRaTER data.

J. G. is supported by the Strategic Priority Program of the Chinese Academy of Sciences (Grant No. XDB41000000 and XDA15017300), the National Natural Science Foundation of China (Grant No. 42074222) and the CNSA pre-research Project on Civil Aerospace Technologies (Grant No. D020104).

M. D. acknowledges support by the EU H2020 Grant Agreement 824135 (SOLARNET) and the Croatian Science Foundation under the Project 7549 (MSOC).

A. P. acknowledges support by the TRACER project (\url{http://members.noa.gr/atpapaio/tracer/}) funded by the National Observatory of Athens (NOA) (Project ID: 5063) and from NASA/LWS project NNH19ZDA001N-LWS.

M. A. and A. B. (IZMIRAN) are supported by the Russian Science Foundation under grant 20-72-10023.

C. M., A. J. W., J. H., T. A. and M. B. thank the Austrian Science Fund (FWF): P31521-N27, P31659-N27, P31265-N27.

This work, as well as the development of EPD on Solar Orbiter were supported by the German Federal Ministry for Economic Affairs and Energy, the German Space Agency (Deutsches Zentrum für Luft- und Raumfahrt e.V., DLR) under grants 50OT0901, 50OT1202, 50OT1702, and 50OT2002, by ESA under contract number SOL.ASTR.CON.00004, the University of Kiel and the Land Schleswig-Holstein, as well as by the Spanish Ministerio de Ciencia, Innovación y Universidades under grants FEDER/MCIU – Agencia Estatal de Investigación/Projects ESP2105-68266-R and ESP2017-88436-R.

The Solar Orbiter magnetometer was funded by the UK Space Agency (grant ST/T001062/1).
	
Solar Orbiter EPD and MAG data are available in the Solar Orbiter Archive at 
\url{http://soar.esac.esa.int/soar/}.

We acknowledge the NMDB database (\url{http://www.nmdb.eu}), funded under the European Union's FP7 Programme (contract 
213007), for providing data. The data from South Pole neutron monitor is provided by the University of Delaware with 
support from the U.S. National Science Foundation under grant ANT‐0838839.

LRO/CRaTER Level 2 data are archived in the NASA Planetary Data System's Planetary Plasma Interactions Node at 
\url{https://pds-ppi.igpp.ucla.edu/} and also available through the CRaTER website at 
\url{https://crater-products.sr.unh.edu/data/inst/l2/}.

The Wind spacecraft solar wind and magnetic field data are provided on the Wind website at \url{https://wind.nasa.gov/mfi_swe_plot.php}.

STEREO heliospheric imager observations and derived data are available on the HELCATS (\url{https://www.helcats-fp7.eu/}) and Helio4Cast (\url{https://helioforecast.space/arrcat}) websites.

Graduated cylindrical shell reconstruction of CMEs was performed using version 0.2.0 of a new Python implementation of the GCS model (\url{https://doi.org/10.5281/zenodo.4443203}) available at \url{https://github.com/johan12345/gcs_python}, which is based on version 2.0.3 (\url{https://doi.org/10.5281/zenodo.4065067}) of the SunPy open source software package \citep{sunpy_community2020} and coronagraph images provided by the Helioviewer.org API \citep{JHelioviewer}.
\end{acknowledgements}


\bibliographystyle{aa}
\bibliography{bibliography}

\begin{thebibliography}{87}
\expandafter\ifx\csname natexlab\endcsname\relax\def\natexlab#1{#1}\fi

\bibitem[{{Abunin} {et~al.}(2013){Abunin}, {Abunina}, {Belov}, {Eroshenko},
  {Oleneva}, {Yanke}, {Mavromichalaki}, \& {Papaioannou}}]{Abunin2013}
{Abunin}, A., {Abunina}, M., {Belov}, A., {et~al.} 2013, in International
  Cosmic Ray Conference, Vol.~33, International Cosmic Ray Conference, 1618

\bibitem[{{Abunina} {et~al.}(2020){Abunina}, {Belov}, {Eroshenko}, {Abunin},
  {Yanke}, {Melkumyan}, {Shlyk}, \& {Pryamushkina}}]{Abunina2020}
{Abunina}, M.~A., {Belov}, A.~V., {Eroshenko}, E.~A., {et~al.} 2020, \solphys,
  295, 69

\bibitem[{Agostinelli {et~al.}(2003)Agostinelli, Allison, Amako, Apostolakis,
  Araujo, Arce, Asai, Axen, Banerjee, Barrand, Behner, Bellagamba, Boudreau,
  Broglia, Brunengo, Burkhardt, Chauvie, Chuma, Chytracek, Cooperman, Cosmo,
  Degtyarenko, Dell'Acqua, Depaola, Dietrich, Enami, Feliciello, Ferguson,
  Fesefeldt, Folger, Foppiano, Forti, Garelli, Giani, Giannitrapani, Gibin,
  {Gómez Cadenas}, González, {Gracia Abril}, Greeniaus, Greiner, Grichine,
  Grossheim, Guatelli, Gumplinger, Hamatsu, Hashimoto, Hasui, Heikkinen,
  Howard, Ivanchenko, Johnson, Jones, Kallenbach, Kanaya, Kawabata, Kawabata,
  Kawaguti, Kelner, Kent, Kimura, Kodama, Kokoulin, Kossov, Kurashige, Lamanna,
  Lampén, Lara, Lefebure, Lei, Liendl, Lockman, Longo, Magni, Maire,
  Medernach, Minamimoto, {Mora de Freitas}, Morita, Murakami, Nagamatu,
  Nartallo, Nieminen, Nishimura, Ohtsubo, Okamura, O'Neale, Oohata, Paech,
  Perl, Pfeiffer, Pia, Ranjard, Rybin, Sadilov, {Di Salvo}, Santin, Sasaki,
  Savvas, Sawada, Scherer, Sei, Sirotenko, Smith, Starkov, Stoecker, Sulkimo,
  Takahata, Tanaka, Tcherniaev, {Safai Tehrani}, Tropeano, Truscott, Uno,
  Urban, Urban, Verderi, Walkden, Wander, Weber, Wellisch, Wenaus, Williams,
  Wright, Yamada, Yoshida, \& Zschiesche}]{Agostinelli-2003}
Agostinelli, S., Allison, J., Amako, K., {et~al.} 2003, Nuclear Instruments and
  Methods in Physics Research Section A: Accelerators, Spectrometers, Detectors
  and Associated Equipment, 506, 250

\bibitem[{Alzate \& Morgan(2017)}]{Alzate_2017}
Alzate, N. \& Morgan, H. 2017, The Astrophysical Journal, 840, 103

\bibitem[{Appel(2018)}]{Appel-2018-PhD}
Appel, J.~K. 2018, PhD thesis, University of Kiel

\bibitem[{Appel {et~al.}(2018)Appel, Köhler, Guo, Ehresmann, Zeitlin,
  Matthiä, Lohf, Wimmer-Schweingruber, Hassler, Brinza, Böhm, Böttcher,
  Martin, Burmeister, Reitz, Rafkin, Posner, Peterson, \& Weigle}]{Appel-2018}
Appel, J.~K., Köhler, J., Guo, J., {et~al.} 2018, Earth and Space Science, 5,
  2

\bibitem[{{Asvestari} {et~al.}(2017){Asvestari}, {Gil}, {Kovaltsov}, \&
  {Usoskin}}]{2017JGRA..122.9790A}
{Asvestari}, E., {Gil}, A., {Kovaltsov}, G.~A., \& {Usoskin}, I.~G. 2017,
  Journal of Geophysical Research (Space Physics), 122, 9790

\bibitem[{Barnes {et~al.}(2019)Barnes, Davies, Harrison, Byrne, Perry, Bothmer,
  Eastwood, Gallagher, Kilpua, M\"{o}stl, Rodriguez, Rouillard, \&
  Odstr{\v{c}}il}]{Barnes-2019}
Barnes, D., Davies, J.~A., Harrison, R.~A., {et~al.} 2019, Solar Physics, 294

\bibitem[{{Belov}(2000)}]{2000SSRv...93...79B}
{Belov}, A. 2000, \ssr, 93, 79

\bibitem[{{Belov} {et~al.}(2015){Belov}, {Abunin}, {Abunina}, {Eroshenko},
  {Oleneva}, {Yanke}, {Papaioannou}, \& {Mavromichalaki}}]{Belov2015}
{Belov}, A., {Abunin}, A., {Abunina}, M., {et~al.} 2015, \solphys, 290, 1429

\bibitem[{{Belov} {et~al.}(2018){Belov}, {Eroshenko}, {Yanke}, {Oleneva},
  {Abunin}, {Abunina}, {Papaioannou}, \& {Mavromichalaki}}]{Belov2018}
{Belov}, A., {Eroshenko}, E., {Yanke}, V., {et~al.} 2018, \solphys, 293, 68

\bibitem[{{Belov} {et~al.}(1974){Belov}, {Blokh}, {Dorman}, {Eroshenko},
  {Inozemtseva}, \& {Kaminer}}]{Belov1974}
{Belov}, A.~V., {Blokh}, I.~L., {Dorman}, L.~I., {et~al.} 1974, Akademiia Nauk
  SSSR Izvestiia Seriia Fizicheskaia, 38, 1867

\bibitem[{{Belov} {et~al.}(1973){Belov}, {Blokh}, {Dorman}, {Eroshenko},
  {Inozemtseva}, \& {Kaminer}}]{Belov1973}
{Belov}, A.~V., {Blokh}, Y.~A., {Dorman}, L.~I., {et~al.} 1973, in
  International Cosmic Ray Conference, Vol.~2, International Cosmic Ray
  Conference, 1247

\bibitem[{Bothmer \& Schwenn(1998)}]{Bothmer-Schwenn-1998}
Bothmer, V. \& Schwenn, R. 1998, Annales Geophysicae, 16, 1

\bibitem[{{Brueckner} {et~al.}(1995){Brueckner}, {Howard}, {Koomen},
  {Korendyke}, {Michels}, {Moses}, {Socker}, {Dere}, {Lamy}, {Llebaria},
  {Bout}, {Schwenn}, {Simnett}, {Bedford}, \& {Eyles}}]{Brueckner-1995-LASCO}
{Brueckner}, G.~E., {Howard}, R.~A., {Koomen}, M.~J., {et~al.} 1995, Solar
  Physics, 162, 357

\bibitem[{Cane(2000)}]{Cane-2000}
Cane, H.~V. 2000, Space Science Reviews, 93, 55

\bibitem[{Cane {et~al.}(1994)Cane, Richardson, von Rosenvinge, \&
  Wibberenz}]{Cane1994}
Cane, H.~V., Richardson, I.~G., von Rosenvinge, T.~T., \& Wibberenz, G. 1994,
  Journal of Geophysical Research: Space Physics, 99, 21429

\bibitem[{Clem \& Dorman(2000)}]{Clem-Dorman-2000}
Clem, J.~M. \& Dorman, L.~I. 2000, Space Science Reviews, 93, 335

\bibitem[{{Corti} {et~al.}(2019){Corti}, {Potgieter}, {Bindi}, {Consolandi},
  {Light}, {Palermo}, \& {Popkow}}]{Corti-2019}
{Corti}, C., {Potgieter}, M.~S., {Bindi}, V., {et~al.} 2019, The Astrophysical
  Journal, 871, 253

\bibitem[{Davies {et~al.}(2021)Davies, M{\"o}stl, Weiss, \& {et
  al.}}]{Davies-2021}
Davies, E., M{\"o}stl, C., Weiss, A.~J., \& {et al.} 2021, Astronomy {\&}
  Astrophysics, submitted (not yet accepted)

\bibitem[{Davies {et~al.}(2012)Davies, Harrison, Perry, Möstl, Lugaz, Rollett,
  Davis, Crothers, Temmer, Eyles, \& Savani}]{Davies-2012-SSE}
Davies, J.~A., Harrison, R.~A., Perry, C.~H., {et~al.} 2012, The Astrophysical
  Journal, 750, 23

\bibitem[{D\'emoulin \& Dasso(2009)}]{Demoulin-Dasso-2009}
D\'emoulin, P. \& Dasso, S. 2009, A\&A, 498, 551

\bibitem[{{D{\'e}moulin} {et~al.}(2008){D{\'e}moulin}, {Nakwacki}, {Dasso}, \&
  {Mandrini}}]{Demoulin-2008}
{D{\'e}moulin}, P., {Nakwacki}, M.~S., {Dasso}, S., \& {Mandrini}, C.~H. 2008,
  Solar Physics, 250, 347

\bibitem[{{Dorman}(2009)}]{Dorman2009}
{Dorman}, L. 2009, {Cosmic Rays in Magnetospheres of the Earth and other
  Planets}, Vol. 358 (Springer)

\bibitem[{{Dumbovi{\'c}} {et~al.}(2018){Dumbovi{\'c}}, Heber, {Vr{\v{s}}nak},
  Temmer, \& Kirin}]{Dumbovic-2018-ForbMod}
{Dumbovi{\'c}}, M., Heber, B., {Vr{\v{s}}nak}, B., Temmer, M., \& Kirin, A.
  2018, The Astrophysical Journal, 860, 71

\bibitem[{{Dumbovi{\'c}} {et~al.}(2020){Dumbovi{\'c}}, {Vr{\v{s}}nak}, {Guo},
  {Heber}, {Dissauer}, {Carcaboso}, {Temmer}, {Veronig}, {Podladchikova},
  {M{\"o}stl}, {Amerstorfer}, \& {Kirin}}]{Dumbovic-2020-ForbMod}
{Dumbovi{\'c}}, M., {Vr{\v{s}}nak}, B., {Guo}, J., {et~al.} 2020, Solar
  Physics, 295, 104

\bibitem[{Forbush(1937)}]{Forbush-1937}
Forbush, S.~E. 1937, Phys. Rev., 51, 1108

\bibitem[{Freiherr~von Forstner {et~al.}(2020)Freiherr~von Forstner, Guo,
  Wimmer-Schweingruber, Dumbovi{\'c}, Janvier, Démoulin, Veronig, Temmer,
  Papaioannou, Dasso, Hassler, \& Zeitlin}]{Forstner-2020}
Freiherr~von Forstner, J.~L., Guo, J., Wimmer-Schweingruber, R.~F., {et~al.}
  2020, Journal of Geophysical Research: Space Physics, 125, e2019JA027662

\bibitem[{Freiherr~von Forstner {et~al.}(2018)Freiherr~von Forstner, Guo,
  Wimmer-Schweingruber, Hassler, Temmer, Dumbovi{\'c}, Jian, Appel,
  Čalogović, Ehresmann, Heber, Lohf, Posner, Steigies, Vršnak, \&
  Zeitlin}]{Forstner-2018}
Freiherr~von Forstner, J.~L., Guo, J., Wimmer-Schweingruber, R.~F., {et~al.}
  2018, Journal of Geophysical Research: Space Physics, 123, 39

\bibitem[{Freiherr~von Forstner {et~al.}(2019)Freiherr~von Forstner, Guo,
  Wimmer-Schweingruber, Temmer, Dumbovi{\'c}, Veronig, Möstl, Hassler,
  Zeitlin, \& Ehresmann}]{Forstner-2019}
Freiherr~von Forstner, J.~L., Guo, J., Wimmer-Schweingruber, R.~F., {et~al.}
  2019, Space Weather, 17, 586

\bibitem[{{Gieseler} \& {Heber}(2016)}]{Gieseler-2016}
{Gieseler}, J. \& {Heber}, B. 2016, Astronomy \& Astrophysics, 589, A32

\bibitem[{{Gieseler} {et~al.}(2017){Gieseler}, {Heber}, \&
  {Herbst}}]{Gieseler-2017}
{Gieseler}, J., {Heber}, B., \& {Herbst}, K. 2017, Journal of Geophysical
  Research (Space Physics), 122, 10,964

\bibitem[{Gopalswamy(2008)}]{Gopalswamy-2008}
Gopalswamy, N. 2008, Journal of Atmospheric and Solar-Terrestrial Physics, 70,
  2078 , coupling of Solar Wind, Magnetosphere, Ionosphere and Upper Atmosphere

\bibitem[{Gulisano {et~al.}(2012)Gulisano, D\'emoulin, Dasso, \&
  Rodriguez}]{Gulisano-2012}
Gulisano, A.~M., D\'emoulin, P., Dasso, S., \& Rodriguez, L. 2012, A\&A, 543,
  A107

\bibitem[{Guo {et~al.}(2020)Guo, Wimmer-Schweingruber, Dumbovi{\'c}, Heber, \&
  Wang}]{Guo-2020-new}
Guo, J., Wimmer-Schweingruber, R.~F., Dumbovi{\'c}, M., Heber, B., \& Wang, Y.
  2020, Earth and Planetary Physics, 4, 62

\bibitem[{Hassler {et~al.}(2012)Hassler, Zeitlin, Wimmer-Schweingruber,
  B{\"o}ttcher, Martin, Andrews, B{\"o}hm, Brinza, Bullock, Burmeister,
  Ehresmann, Epperly, Grinspoon, K{\"o}hler, Kortmann, Neal, Peterson, Posner,
  Rafkin, Seimetz, Smith, Tyler, Weigle, Reitz, \&
  Cucinotta}]{Hassler-2012-MSLRAD}
Hassler, D.~M., Zeitlin, C., Wimmer-Schweingruber, R.~F., {et~al.} 2012, Space
  Science Reviews, 170, 503

\bibitem[{He {et~al.}(2018)He, Liu, Hu, Wang, \& Zhao}]{He-2018}
He, W., Liu, Y.~D., Hu, H., Wang, R., \& Zhao, X. 2018, The Astrophysical
  Journal, 860, 78

\bibitem[{Horbury {et~al.}(2020)Horbury, O'Brien, Blazquez, Bendyk, Brown,
  Hudson, Evans, Carr, Beek, Bhattacharya, Dominguez, Matthews, Myklebust,
  Whiteside, Bale, Baumjohann, Bavassano, \& Burgess}]{Horbury-2020-MAG}
Horbury, T.~S., O'Brien, H., Blazquez, I.~C., {et~al.} 2020, Astronomy {\&}
  Astrophysics

\bibitem[{Howard {et~al.}(2008)Howard, Moses, Vourlidas, Newmark, Socker,
  Plunkett, Korendyke, Cook, Hurley, Davila, Thompson, Cyr, Mentzell, Mehalick,
  Lemen, Wuelser, Duncan, Tarbell, Wolfson, Moore, Harrison, Waltham, Lang,
  Davis, Eyles, Mapson-Menard, Simnett, Halain, Defise, Mazy, Rochus, Mercier,
  Ravet, Delmotte, Auchere, Delaboudiniere, Bothmer, Deutsch, Wang, Rich,
  Cooper, Stephens, Maahs, Baugh, McMullin, \& Carter}]{Howard-2008-SECCHI}
Howard, R.~A., Moses, J.~D., Vourlidas, A., {et~al.} 2008, Space Science
  Reviews, 136, 67

\bibitem[{{Howard} \& {Harrison}(2013)}]{Howard-2013-StealthCMEs}
{Howard}, T.~A. \& {Harrison}, R.~A. 2013, Solar Physics, 285, 269

\bibitem[{Janvier {et~al.}(2019)Janvier, Winslow, Good, Bonhomme, Démoulin,
  Dasso, Möstl, Lugaz, Amerstorfer, Soubrié, \& Boakes}]{Janvier-2019}
Janvier, M., Winslow, R.~M., Good, S., {et~al.} 2019, Journal of Geophysical
  Research: Space Physics, 124, 812

\bibitem[{Kilpua {et~al.}(2017)Kilpua, Koskinen, \& Pulkkinen}]{Kilpua2017}
Kilpua, E., Koskinen, H. E.~J., \& Pulkkinen, T.~I. 2017, Living Reviews in
  Solar Physics, 14, 5

\bibitem[{{Koldobskiy} {et~al.}(2018){Koldobskiy}, {Kovaltsov}, \&
  {Usoskin}}]{2018SoPh..293..110K}
{Koldobskiy}, S.~A., {Kovaltsov}, G.~A., \& {Usoskin}, I.~G. 2018, \solphys,
  293, 110

\bibitem[{Krymsky(1964)}]{krymsky1964diffusion}
Krymsky, G. 1964, Geomagn. Aeronomy, 4, 763

\bibitem[{Krymsky {et~al.}(1966)Krymsky, Altukhov, Kuzmin, \&
  Skripin}]{krymsky1966new}
Krymsky, G., Altukhov, A., Kuzmin, A., \& Skripin, G. 1966, A New Method for
  Studying the Anisotropy of Cosmic Rays--Investigation of Geomagnetism and
  Aeronomy (Nauka, Moscow)

\bibitem[{{K{\"u}hl} {et~al.}(2015){K{\"u}hl}, {Banjac}, {Heber}, {Labrenz},
  {M{\"u}ller-Mellin}, \& {Terasa}}]{Kuehl-2015}
{K{\"u}hl}, P., {Banjac}, S., {Heber}, B., {et~al.} 2015, Central European
  Astrophysical Bulletin, 39, 119

\bibitem[{{Lawrence} {et~al.}(2016){Lawrence}, {Peplowski}, {Feldman},
  {Schwadron}, \& {Spence}}]{Lawrence-2016}
{Lawrence}, D.~J., {Peplowski}, P.~N., {Feldman}, W.~C., {Schwadron}, N.~A., \&
  {Spence}, H.~E. 2016, Journal of Geophysical Research (Space Physics), 121,
  7398

\bibitem[{Leitner {et~al.}(2007)Leitner, Farrugia, Möstl, Ogilvie, Galvin,
  Schwenn, \& Biernat}]{Leitner-2007}
Leitner, M., Farrugia, C.~J., Möstl, C., {et~al.} 2007, Journal of Geophysical
  Research: Space Physics, 112

\bibitem[{{Lemen} {et~al.}(2012){Lemen}, {Title}, {Akin}, {Boerner}, {Chou},
  {Drake}, {Duncan}, {Edwards}, {Friedlaender}, {Heyman}, {Hurlburt}, {Katz},
  {Kushner}, {Levay}, {Lindgren}, {Mathur}, {McFeaters}, {Mitchell}, {Rehse},
  {Schrijver}, {Springer}, {Stern}, {Tarbell}, {Wuelser}, {Wolfson}, {Yanari},
  {Bookbinder}, {Cheimets}, {Caldwell}, {Deluca}, {Gates}, {Golub}, {Park},
  {Podgorski}, {Bush}, {Scherrer}, {Gummin}, {Smith}, {Auker}, {Jerram},
  {Pool}, {Soufli}, {Windt}, {Beardsley}, {Clapp}, {Lang}, \&
  {Waltham}}]{Lemen-2012-AIA}
{Lemen}, J.~R., {Title}, A.~M., {Akin}, D.~J., {et~al.} 2012, Solar Physics,
  275, 17

\bibitem[{Lepping {et~al.}(1995)Lepping, Ac{\~{u}}na, Burlaga, Farrell, Slavin,
  Schatten, Mariani, Ness, Neubauer, Whang, Byrnes, Kennon, Panetta, Scheifele,
  \& Worley}]{Lepping-1995-WindMFI}
Lepping, R.~P., Ac{\~{u}}na, M.~H., Burlaga, L.~F., {et~al.} 1995, Space
  Science Reviews, 71, 207

\bibitem[{Lockwood(1971)}]{Lockwood1971}
Lockwood, J.~A. 1971, Space Science Reviews, 12, 658

\bibitem[{Lockwood {et~al.}(1991)Lockwood, Webber, \& Debrunner}]{Lockwood1991}
Lockwood, J.~A., Webber, W.~R., \& Debrunner, H. 1991, Journal of Geophysical
  Research: Space Physics, 96, 5447

\bibitem[{Looper {et~al.}(2013)Looper, Mazur, Blake, Spence, Schwadron,
  Golightly, Case, Kasper, \& Townsend}]{Looper-2013}
Looper, M.~D., Mazur, J.~E., Blake, J.~B., {et~al.} 2013, Space Weather, 11,
  142

\bibitem[{Ma {et~al.}(2010)Ma, Attrill, Golub, \& Lin}]{Ma-2010}
Ma, S., Attrill, G. D.~R., Golub, L., \& Lin, J. 2010, The Astrophysical
  Journal, 722, 289

\bibitem[{Manchester {et~al.}(2005)Manchester, Gombosi, Zeeuw, Sokolov,
  Roussev, Powell, Kota, Toth, \& Zurbuchen}]{Manchester2005}
Manchester, IV, W.~B., Gombosi, T.~I., Zeeuw, D. L.~D., {et~al.} 2005, The
  Astrophysical Journal, 622, 1225

\bibitem[{{McDonald} \& {Ludwig}(1964)}]{McDonald-1964}
{McDonald}, F.~B. \& {Ludwig}, G.~H. 1964, Physical Review Letters, 13, 783

\bibitem[{M\"{u}ller {et~al.}(2020)M\"{u}ller, Cyr, Zouganelis, Gilbert,
  Marsden, \& and}]{Mueller-2020-SolO}
M\"{u}ller, D., Cyr, O.~S., Zouganelis, I., {et~al.} 2020, Astronomy {\&}
  Astrophysics

\bibitem[{{M{\"u}ller} {et~al.}(2013){M{\"u}ller}, {Marsden}, {St. Cyr}, \&
  {Gilbert}}]{Mueller-2013-SolO}
{M{\"u}ller}, D., {Marsden}, R.~G., {St. Cyr}, O.~C., \& {Gilbert}, H.~R. 2013,
  Solar Physics, 285, 25

\bibitem[{M\"uller {et~al.}(2017)M\"uller, Nicula, Felix, Verstringe,
  Bourgoignie, Csillaghy, Berghmans, Jiggens, Garc\'{\i}a-Ortiz, Ireland,
  Zahniy, \& Fleck}]{JHelioviewer}
M\"uller, D., Nicula, B., Felix, S., {et~al.} 2017, A\&A, 606, A10

\bibitem[{Möstl {et~al.}(2018)Möstl, Amerstorfer, Palmerio, Isavnin,
  Farrugia, Lowder, Winslow, Donnerer, Kilpua, \& Boakes}]{Moestl2018-3DCORE}
Möstl, C., Amerstorfer, T., Palmerio, E., {et~al.} 2018, Space Weather, 16,
  216

\bibitem[{Möstl {et~al.}(2017)Möstl, Isavnin, Boakes, Kilpua, Davies,
  Harrison, Barnes, Krupar, Eastwood, Good, Forsyth, Bothmer, Reiss,
  Amerstorfer, Winslow, Anderson, Philpott, Rodriguez, Rouillard, Gallagher,
  Nieves‐Chinchilla, \& Zhang}]{Moestl-2017-HelcatsHSO}
Möstl, C., Isavnin, A., Boakes, P.~D., {et~al.} 2017, Space Weather, 15, 955

\bibitem[{Ogilvie {et~al.}(1995)Ogilvie, Chornay, Fritzenreiter, Hunsaker,
  Keller, Lobell, Miller, Scudder, Sittler, Torbert, Bodet, Needell, Lazarus,
  Steinberg, Tappan, Mavretic, \& Gergin}]{Ogilvie-1995-WindSWE}
Ogilvie, K.~W., Chornay, D.~J., Fritzenreiter, R.~J., {et~al.} 1995, Space
  Science Reviews, 71, 55

\bibitem[{Owen {et~al.}(2020)Owen, Bruno, Livi, Louarn, \&
  Janabi}]{Owen-2020-SWA}
Owen, C.~J., Bruno, R., Livi, S., Louarn, P., \& Janabi, K.~A. 2020, Astronomy
  {\&} Astrophysics

\bibitem[{{Papaioannou} {et~al.}(2020){Papaioannou}, {Belov}, {Abunina},
  {Eroshenko}, {Abunin}, {Anastasiadis}, {Patsourakos}, \&
  {Mavromichalaki}}]{Papaioannou2020}
{Papaioannou}, A., {Belov}, A., {Abunina}, M., {et~al.} 2020, \apj, 890, 101

\bibitem[{{Papaioannou} {et~al.}(2019){Papaioannou}, {Belov}, {Abunina}, {Guo},
  {Anastasiadis}, {Wimmer-Schweingruber}, {Eroshenko}, {Melkumyan}, {Abunin},
  {Heber}, {Herbst}, \& {Steigies}}]{Papaioannou2019}
{Papaioannou}, A., {Belov}, A., {Abunina}, M., {et~al.} 2019, \solphys, 294, 66

\bibitem[{{Potgieter}(2013)}]{Potgieter-2013}
{Potgieter}, M.~S. 2013, Living Reviews in Solar Physics, 10, 3

\bibitem[{{Potgieter} {et~al.}(2014){Potgieter}, {Vos}, {Boezio}, {De Simone},
  {Di Felice}, \& {Formato}}]{Potgieter-2014}
{Potgieter}, M.~S., {Vos}, E.~E., {Boezio}, M., {et~al.} 2014, Solar Physics,
  289, 391

\bibitem[{Reames(2013)}]{Reames-2013}
Reames, D.~V. 2013, Space Science Reviews, 175, 53

\bibitem[{{Richardson} \& {Cane}(2011)}]{2011SoPh..270..609R}
{Richardson}, I.~G. \& {Cane}, H.~V. 2011, \solphys, 270, 609

\bibitem[{Richardson {et~al.}(1996)Richardson, Wibberenz, \&
  Cane}]{Richardson-1996}
Richardson, I.~G., Wibberenz, G., \& Cane, H.~V. 1996, Journal of Geophysical
  Research: Space Physics, 101, 13483

\bibitem[{Rodr\'{\i}guez-Pacheco {et~al.}(2020)Rodr\'{\i}guez-Pacheco,
  Wimmer-Schweingruber, Mason, Ho, {S\'anchez-Prieto, S.}, {Prieto, M.},
  {Mart\'{\i}n, C.}, {Seifert, H.}, {Andrews, G. B.}, {Kulkarni, S. R.},
  {Panitzsch, L.}, {Boden, S.}, {B\"ottcher, S. I.}, {Cernuda, I.}, {Elftmann,
  R.}, {Espinosa Lara, F.}, {G\'omez-Herrero, R.}, {Terasa, C.}, {Almena, J.},
  {Begley, S.}, {B\"ohm, E.}, {Blanco, J. J.}, {Boogaerts, W.}, {Carrasco, A.},
  {Castillo, R.}, {da Silva Fari\~na, A.}, {de Manuel Gonz\'alez, V.}, {Drews,
  C.}, {Dupont, A. R.}, {Eldrum, S.}, {Gordillo, C.}, {Guti\'errez, O.},
  {Haggerty, D. K.}, {Hayes, J. R.}, {Heber, B.}, {Hill, M. E.}, {J\"ungling,
  M.}, {Kerem, S.}, {Knierim, V.}, {K\"ohler, J.}, {Kolbe, S.}, {Kulemzin, A.},
  {Lario, D.}, {Lees, W. J.}, {Liang, S.}, {Mart\'{\i}nez Hell\'{\i}n, A.},
  {Meziat, D.}, {Montalvo, A.}, {Nelson, K. S.}, {Parra, P.}, {Paspirgilis,
  R.}, {Ravanbakhsh, A.}, {Richards, M.}, {Rodr\'{\i}guez-Polo, O.}, {Russu,
  A.}, {S\'anchez, I.}, {Schlemm, C. E.}, {Schuster, B.}, {Seimetz, L.},
  {Steinhagen, J.}, {Tammen, J.}, {Tyagi, K.}, {Varela, T.}, {Yedla, M.}, {Yu,
  J.}, {Agueda, N.}, {Aran, A.}, {Horbury, T. S.}, {Klecker, B.}, {Klein,
  K.-L.}, {Kontar, E.}, {Krucker, S.}, {Maksimovic, M.}, {Malandraki, O.},
  {Owen, C. J.}, {Pacheco, D.}, {Sanahuja, B.}, {Vainio, R.}, {Connell, J. J.},
  {Dalla, S.}, {Dr\"oge, W.}, {Gevin, O.}, {Gopalswamy, N.}, {Kartavykh, Y.
  Y.}, {Kudela, K.}, {Limousin, O.}, {Makela, P.}, {Mann, G.}, {\"Onel, H.},
  {Posner, A.}, {Ryan, J. M.}, {Soucek, J.}, {Hofmeister, S.}, {Vilmer, N.},
  {Walsh, A. P.}, {Wang, L.}, {Wiedenbeck, M. E.}, {Wirth, K.}, \& {Zong,
  Q.}}]{RodriguezPacheco-2019-EPD}
Rodr\'{\i}guez-Pacheco, J., Wimmer-Schweingruber, R.~F., Mason, G.~M., {et~al.}
  2020, A\&A, 642, A7

\bibitem[{Schwadron {et~al.}(2012)Schwadron, Baker, Blake, Case, Cooper,
  Golightly, Jordan, Joyce, Kasper, Kozarev, Mislinski, Mazur, Posner, Rother,
  Smith, Spence, Townsend, Wilson, \& Zeitlin}]{Schwadron-2012}
Schwadron, N.~A., Baker, T., Blake, B., {et~al.} 2012, Journal of Geophysical
  Research: Planets, 117

\bibitem[{{Simpson}(1983)}]{Simpson-1983}
{Simpson}, J.~A. 1983, Annual Review of Nuclear and Particle Science, 33, 323

\bibitem[{Siscoe \& Odstrcil(2008)}]{Siscoe2008}
Siscoe, G. \& Odstrcil, D. 2008, Journal of Geophysical Research: Space
  Physics, 113

\bibitem[{Smart \& Shea(2008)}]{Smart-Shea-2008}
Smart, D. \& Shea, M. 2008, in Proceedings of the 30th International Cosmic Ray
  Conference, Vol.~1, 737--740

\bibitem[{{Sohn} {et~al.}(2019{\natexlab{a}}){Sohn}, {Oh}, {Yi}, \&
  {Lee}}]{Sohn-2019-Enhancement}
{Sohn}, J., {Oh}, S., {Yi}, Y., \& {Lee}, J. 2019{\natexlab{a}}, Journal of
  Korean Physical Society, 74, 614

\bibitem[{{Sohn} {et~al.}(2019{\natexlab{b}}){Sohn}, {Oh}, {Yi}, \&
  {Lee}}]{Sohn-2019-Forbush}
{Sohn}, J., {Oh}, S., {Yi}, Y., \& {Lee}, J. 2019{\natexlab{b}}, Astrophysics
  and Space Science, 364, 125

\bibitem[{{Spence} {et~al.}(2010){Spence}, {Case}, {Golightly}, {Heine},
  {Larsen}, {Blake}, {Caranza}, {Crain}, {George}, {Lalic}, {Lin}, {Looper},
  {Mazur}, {Salvaggio}, {Kasper}, {Stubbs}, {Doucette}, {Ford}, {Foster},
  {Goeke}, {Gordon}, {Klatt}, {O'Connor}, {Smith}, {Onsager}, {Zeitlin},
  {Townsend}, \& {Charara}}]{Spence-2010}
{Spence}, H.~E., {Case}, A.~W., {Golightly}, M.~J., {et~al.} 2010, Space
  Science Reviews, 150, 243

\bibitem[{{The SunPy Community} {et~al.}(2020){The SunPy Community}, Barnes,
  Bobra, Christe, Freij, Hayes, Ireland, Mumford, Perez-Suarez, Ryan, Shih,
  Chanda, Glogowski, Hewett, Hughitt, Hill, Hiware, Inglis, Kirk, Konge, Mason,
  Maloney, Murray, Panda, Park, Pereira, Reardon, Savage, Sipőcz, Stansby,
  Jain, Taylor, Yadav, Rajul, \& Dang}]{sunpy_community2020}
{The SunPy Community}, Barnes, W.~T., Bobra, M.~G., {et~al.} 2020, The
  Astrophysical Journal, 890, 68

\bibitem[{Thernisien(2011)}]{Thernisien-2011-GCS}
Thernisien, A. 2011, The Astrophysical Journal Supplement Series, 194, 33

\bibitem[{Thernisien {et~al.}(2006)Thernisien, Howard, \&
  Vourlidas}]{Thernisien-2006-GCS}
Thernisien, A., Howard, R.~A., \& Vourlidas, A. 2006, The Astrophysical
  Journal, 652, 763

\bibitem[{{Usoskin} {et~al.}(2011){Usoskin}, {Bazilevskaya}, \&
  {Kovaltsov}}]{Usoskin-2011}
{Usoskin}, I.~G., {Bazilevskaya}, G.~A., \& {Kovaltsov}, G.~A. 2011, Journal of
  Geophysical Research (Space Physics), 116, A02104

\bibitem[{{Webber} \& {Lockwood}(1999)}]{Webber-1999}
{Webber}, W.~R. \& {Lockwood}, J.~A. 1999, Journal of Geophysical Research,
  104, 2487

\bibitem[{{Weiss} {et~al.}(2020){Weiss}, {M{\"o}stl}, {Amerstorfer}, {Bailey},
  {Reiss}, {Hinterreiter}, {Amerstorfer}, \& {Bauer}}]{Weiss2020-3DCORE}
{Weiss}, A.~J., {M{\"o}stl}, C., {Amerstorfer}, T., {et~al.} 2020, arXiv
  e-prints, arXiv:2009.00327

\bibitem[{Wimmer-Schweingruber {et~al.}(2020)Wimmer-Schweingruber, Yu,
  Böttcher, Zhang, Burmeister, Lohf, Guo, Xu, Schuster, Seimetz, von Forstner,
  Ravanbakhsh, Knierim, Kolbe, Woyciechowsky, Kulkarni, Yuan, Shen, Wang,
  Chang, Berger, Hellweg, Matthiä, Hou, Knappmann, Büschel, Hou, Ren, \&
  Fu}]{Wimmer-2020-LND}
Wimmer-Schweingruber, R.~F., Yu, J., Böttcher, S.~I., {et~al.} 2020, Space
  Science Reviews, 216

\bibitem[{Winslow {et~al.}(2018)Winslow, Schwadron, Lugaz, Guo, Joyce, Jordan,
  Wilson, Spence, Lawrence, Wimmer-Schweingruber, \& Mays}]{Winslow-2018}
Winslow, R.~M., Schwadron, N.~A., Lugaz, N., {et~al.} 2018, The Astrophysical
  Journal, 856, 139

\bibitem[{Witasse {et~al.}(2017)Witasse, Sánchez-Cano, Mays, Kajdič,
  Opgenoorth, Elliott, Richardson, Zouganelis, Zender, Wimmer-Schweingruber,
  Turc, Taylor, Roussos, Rouillard, Richter, Richardson, Ramstad, Provan,
  Posner, Plaut, Odstrcil, Nilsson, Niemenen, Milan, Mandt, Lohf, Lester,
  Lebreton, Kuulkers, Krupp, Koenders, James, Intzekara, Holmstrom, Hassler,
  Hall, Guo, Goldstein, Goetz, Glassmeier, Génot, Evans, Espley, Edberg,
  Dougherty, Cowley, Burch, Behar, Barabash, Andrews, \&
  Altobelli}]{Witasse-2017}
Witasse, O., Sánchez-Cano, B., Mays, M.~L., {et~al.} 2017, Journal of
  Geophysical Research: Space Physics, 122, 7865, 2017JA023884

\end{thebibliography}

\end{document}